\documentclass{jfm}
\usepackage{amsmath}
\usepackage{graphicx}
\usepackage{graphics}
\usepackage{epstopdf}
\usepackage{subcaption}
\usepackage{natbib}
\usepackage{apalike}
\usepackage{color}
\usepackage[dvipsnames]{xcolor}

\shorttitle{Transient energy growth in the ageostrophic Eady model}
\shortauthor{V. E. Zemskova,  P.-Y. Passaggia, B. L. White}

\title{Transient energy growth in the ageostrophic Eady model}

\author{Varvara E. Zemskova$^1$
  \corresp{\email{zemskova@live.unc.edu}},
  Pierre-Yves Passaggia$^{1,2}$
 \and Brian L. White$^1$}

\affiliation{$^1$Department of Marine Sciences, University of North Carolina, Chapel Hill, NC 27599, U.S.A.\\
$^2$ University of Orl{\'e}ans, INSA-CVL, PRISME, EA 4229, 45072, Orl{\'e}ans, France
}

\begin{document}

\maketitle

\begin{abstract}

The problem of optimal initial disturbances in thermal wind shear is revisited and extended to include non-hydrostatic effects. 
This systematic study compares transient and modal growth rates of submesoscale instabilities over a large range of zonal and meridional wave numbers, aspect ratios, and different Richardson number regimes. 
Selection criteria were derived to remove spurious and unresolved instability modes that arise from the eigenvalue problem and we generalize the study of the hydrostatic Eady problem by \citet{Heifetz:03,heifetz2007generalized,heifetz2008non} to thin fronts, characterized by large aspect ratios. Such fronts are commonly found at the early stages of frontogenesis, for example, in the ocean mesoscale eddies and near the eye wall of hurricanes.
In particular, we show that transient energy growth rates are up to two orders of magnitude larger than modal counterparts for a wide range of Richardson number and that the effects of transient energy gain become even greater when non-hydrostatic effects become important and/or for large Richardson numbers.
This study also compares the dominant energy pathways contributing to the energy growth at short and long times. 
For symmetric modes,
we recover the inertia-gravity instability described in \citet{Xu:07}. 
These mechanisms are shown to be the most powerful mediator of vertical transport when compared with the fastest growing baroclinic and symmetric modes.
These results highlight the importance of transient processes in the ocean and the atmosphere.
\end{abstract}

\section{Introduction}

Density fronts, which are defined as regions of large density gradients, are features ubiquitous in the ocean and the atmosphere. For instance western boundary currents, such as the Gulf Stream and the Kuroshio, are the site of large-scale meanders resulting in mesoscale eddies, which are typically $10-100$ km in horizontal length scale with Rossby numbers 
smaller
than unity. These eddies draw energy from the geostrophic flow by converting potential into kinetic energy \citep{Gnanadesikan:05,Wolfe:08,Thomas:13,Zemskova:15}. The kinetic energy in these eddies can in turn dissipate through submesoscale instabilities (typically smaller than $1$ km), in addition to other mechanisms such as bottom drag, but the transient dynamics of the submesoscale instabilities  are still not completely understood \citep{Thomas:13,Stamper:17}.
These submesoscale processes enhance the vertical transport in the upper ocean and have an impact on phytoplankton and the biological pump, and their energetics need to be explored further
\citep{Taylor:11,Omand:15,sarkar2016interplay,brannigan2017submesoscale,ramachandran2018submesoscale}. \\

Small-scale perturbations have also been identified as an energy source for geophysical shear flows with undisturbed background state, such as the Earth's midlatitude jets \citep{farrell1993stochastic_b} and zonal winds in gaseous planets \citep{vasavada2005jovian}, with a particular interest on the effect of these perturbations on cyclogenesis.
However, the stability and dynamics of the fronts where small-scale instabilities are confined in narrow regions have received less attention than the small aspect ratio and hydrostatic counterpart.
Larger aspect ratios are particularly common in hurricane boundary layers \citep{ellis2010helical,foster2013signature} and tornados \citep{nolan2017tornado}.
This confinement reduces the spatial support for large baroclinic geostrophic instabilities where non-hydrostatic effects are known to damp geostrophic-type motions \citep{Stone:70}. Hence ageostrophic-type modes, characterized by short wavelength and critical regions associated with intense shear, are expected to play a dominant role in the submesoscale transition process. \\

The role of submesoscale instabilities was first identified by \citet{Solberg:36} in the case of symmetric modes
and later by \citet{Eady:49} for baroclinic modes and finally by \citet{Chandrasekhar:61} for Kelvin-Helmholtz type waves. The stability of fronts and normal growth rates of instabilities were extensively analyzed using asymptotic theory for a linear thermal wind shear \citep{Stone:66,Stone:70,Stone:71}. In general, baroclinic instabilities include any instabilities that arise in a rotating fluid which is stratified due to horizontal density gradients. Here, to be consistent with Stone's work and subsequent analyses done in his footsteps \cite[e.g.][]{MolemakerM:05,Stamper:17}, we refer to baroclinic instabilities as those that occur independently of perturbations in the meridional direction and draw energy from potential energy of the flow to distinguish them from 
symmetric modes
that are independent of zonal perturbations and draw energy from the kinetic energy field.\\

A geostrophic current is symmetrically unstable  when its potential vorticity has the opposite sign of local Coriolis parameter \citep{Thomas:13}, but the unstable modal growth rate is nonzero only when $\mbox{Ri} < 1$ for a basic state in thermal wind shear balance with no vertical vorticity \citep{Stone:66}. Baroclinic instabilities are divided into geostrophic and ageostrophic types. While these instabilities are technically geostrophic only for $\mbox{Ri} \gg 1$ as described by \cite{Eady:49}, we again follow Stone's convention in referring to any baroclinic instabilities at any value of $\mbox{Ri}$ with large wavelength and large modal growth rates as geostrophic, while the ageostrophic ones occur at much smaller length scales and have smaller growth rates. The geostrophic and ageostrophic instabilities also have different structures of the eigenspectra, as will be discussed in detail in \S 2.3 and \S 4.1.  In particular, the geostrophic-type instabilities have only one eigenmode with the largest real part of eigenvalue while the ageostrophic-type instabilities have two eigenmodes with the largest equal real part but different imaginary parts of the eigenvalue. Ageostrophic modes have been previously overlooked, but it has been shown that they may play an important role in restratification of ocean mixed layer \citep{Boccaletti:07} and loss of balance \citep{MolemakerM:05}. 
\\

\citet{Heifetz:03} generalized the notion of stability in baroclinic shear flows in the large Richardson (strongly stratified) regime using a formulation based on primitive equations. 
In a second study, \citet{heifetz2007generalized} showed that the non-normal coupling between stable (Rossby- or inertio-gravity) waves and unstable geostrophic modes results in energy gain which exceeds that provided by the modal growth alone in the hydrostatic limit. \citet{heifetz2008non} also extended their analysis to 
symmetric modes
and showed the optimal transient growth exceeding that predicted by normal symmetric mode analysis and yielding potentially a much faster generation of slantwise convection. However, the extra gains that they report are at most a factor two, which is relatively small compared to the extra optimal gain observed for the geostrophic and ageostrophic type modes. 
For all cases, this shear-driven transient energy growth is known to be the bypass transition to turbulence, which in turn was developed by the hydrodynamic community for asymptotically stable shear flows by \cite{schmid2012stability}. Finite amplitude perturbations that undergo transient growth may reach an amplitude that is sufficiently large to allow positive feedback through nonlinear interactions that amplify the growing disturbances. 
In the present study, we consider a generalization of \citet{Heifetz:03,heifetz2007generalized} by taking it to the non-hydrostatic limit commonly found during frontogenesis that induces finite thicknesses of the front leading to geostrophic shear with large aspect ratios.\\

While oceanic flows are generally characterized by small aspect ratios (defined here as the ratio of the vertical to horizontal length scales), commonly found  between $10^{-1}$ and $10^{-3}$, atmospheric flows can have aspect ratios of the order of unity \citep{nolan2017tornado} and the nonhydrostatic effects may be important.
\cite{Stone:71} theoretically derived the nonhydrostatic effects on the most unstable modes for the linear thermal wind shear framework. This work found that the temporal energy growth rates are progressively suppressed as the aspect ratio increases for all instability types. However, it is possible that for a short-term, the energy gain is not as affected by non-hydrostatic conditions. The latter may even provide a route for the transient growth of perturbations
which will be one of the main focus of this study. For instance, the energy gain of short-lived near-surface streaks in the hurricane boundary layer that enhance wind speed and vertical transport could be attributed to such non-normal transient growth dynamics \citep{drobinski2003origin}. \\

Unstratified simple shear flows such as plane Couette or plane Poiseuille configurations are known to be subject to two types of transient growth phenomena. In his original work, \citet{Orr:07} showed that in the case of a simple inviscid parallel shear flow, perturbations with a non-zero streamwise wavenumber could produce transient growth through the kinematic deformation of the perturbation vorticity by the baseflow advection and shear.
Later, \citet{farrell1993optimal} derived an analytic solution for the Orr temporal growth rate, for two-dimensional perturbations in a constant, unstratified linear shear. 
\citet{ellingsen1975stability} recognized that a finite disturbance independent of the streamwise coordinate (i.e. with a spanwise wavenumber with respect to the shear) may lead to instability of linear flow, even though the basic velocity does not possesses an inflection point. This mechanism, later denoted as lift-up effect, is a key process in the laminar-turbulent transition in shear flows and in fully developed turbulence \citep{brandt2014lift}.  \\

The energy growth mechanisms in stratified and rotating sheared flows have been addressed in several previous works, such as \citet{farrell1993stochastic_a,bakas2009gravity_a,bakas2009gravity_b, park2017instabilities}. These works identified several mechanisms responsible for the energy growth that vary depending on the configuration of the background flow and the stratification. In particular, \citet{park2017instabilities} noted that there exist two mechanisms types for transient energy growth analogous to the lift up and the Orr mechanisms in homogeneous fluids but with the additional effect of density perturbations. 
In this work, we explore the transient energy growth mechanisms for different Richardson number regimes over a range of zonal and meridional wavenumbers with respect to the height or depth of the front.  In particular, we assess the additional contribution of the energy transfer from the background flow to the perturbations through the wave reflection off the domain boundaries and critical layers.\\

The local stability properties and loss of balance induced by baroclinic modes and their dependence on Rossby numbers in thermal wind balance was first computed numerically using a finite difference matrix-type approach to this eigenvalue problem rather than an iterative shooting method with an initial eigenvalue guess by \citet{MolemakerM:05}. They showed that they could capture the instability modes predicted by \citet{Stone:70} and that these modes explained in part how a highly balanced large-scale circulation
may dissipate energy through a local forward energy cascade into unbalanced motions. \citet{Nakamura:88} showed that ageostrophic modes are identified within the inertial critical layer which is sustained by the resonance between one of the boundary modes and the inertial gravity waves. 
The latter observation implies that computing the eigenspectrum in the non-hydrostatic and inviscid limit based on \citet{MolemakerM:05} formulation may lead to spurious or unresolved eigenfunctions in the critical layers. The computation of optimal initial disturbances and optimal energy gain using a singular value decomposition method described by \cite{Schmid:14} requires the interaction of all modes, not only modes with non-zero energy growth rate. As some of the obtained eigenfunctions may be spurious or unresolved, it is necessary to determine the criteria for the selection of the only physical and resolved eigenmodes appropriate for the these computations.\\

In the present paper, we start from a one-dimensional finite difference approximation of the linearized Euler equation for a flow in thermal wind balance following \citet{Stone:66,Stone:70,Stone:71}, and compute optimal energy growth for a range of Richardson number values as well as wavenumbers in both horizontal directions. We overcome the problem of critical layers by performing a careful selection of the appropriate modes used in the Galerkin-type projection to compute the optimal initial disturbance. In particular, we provide selection criteria that allow for discarding the spurious eigenfunctions. The energy budget and nonlinear development of the optimal initial conditions are computed using three-dimensional non-hydrostatic direct numerical simulations (DNS) of the Navier-Stokes equations.\\

The rest of the paper is organized as follows: \S 2 presents the numerical methods and setup of the problem, transient growth and the selection criteria for the eigenfunctions are explained in \S 3. Results for the linear optimal dynamics are summarized in \S 4, whereas three-dimensional simulations are given in \S 5 and conclusions are drawn in \S 6.

\section{Theoretical formulation}\label{sec:theory}

\subsection{Problem set-up}
The present formulation is based on \citet{Stone:70,Stone:71} and considers the dynamics of a non-hydrostatic and rotating inviscid fluid in the Boussinesq limit where $f$ is the Coriolis frequency. The governing conservation equations for the zonal velocity $\hat{u}$, meridional velocity $\hat{v}$, vertical velocity $\hat{w}$, hydrodynamic pressure $\hat{p}' = \hat{p}/\rho_0$, and buoyancy $\hat{b} = -g (\hat{\rho}-\hat{\rho_0})/\hat{\rho_0}$ (where $g$ is the gravitational acceleration) are
\begin{equation}
\frac{D \hat{\mathbf{u}}}{D t} + f\mathbf{e_z}\times\hat{\mathbf{u}} = -\nabla \hat{p}' + \hat{b}\mathbf{e_z}, \, \, \, \nabla\cdot\hat{\mathbf{u}}=0, \, \, \,\frac{D \hat{b}}{D t}=0,
\label{Euler}
\end{equation}
where $D/D\hat{t} = \partial/\partial \hat{t} + \hat{\mathbf{u}}\cdot\nabla$, and $  \mathbf{e_z}$ is the vertical unit vector positive upward.
The domain is unbounded in $\hat{x}$ (zonal) and $\hat{y}$ (meridional) directions with solid vertical boundaries at $\hat{z}=[0,H]$, where we impose the boundary condition $\hat{w}=0$. The base buoyancy field has constant vertical stratification and gradient in the meridional direction. The base state has zero meridional and vertical velocities and the zonal velocity has a constant vertical shear in a thermal wind balance \citep{vallis}. 

The governing equations (\ref{Euler}) are non-dimensionalised following the time and length scales prescribed in \citet{Stone:71} such that
\begin{equation}
\begin{split}
(x,y)=\frac{f(\hat{x},\hat{y})}{u_0}, \, \, \, z=\frac{\hat{z}}{H}, \, \, \, t=\hat{t}f, \\
(u,v)=\frac{(\hat{u},\hat{v})}{u_0}, \, \, \, w=\frac{\hat{w}}{fH},  \, \, \, p=\frac{\hat{p}'}{N^2 H^2},  \, \, \, b=\frac{\hat{b}}{H N^2},
\end{split}
\label{adim}
\end{equation}
where $u_0$ is the maximum zonal velocity and $N=\partial b/\partial z$ is the Brunt-V\"ais\"al\"a frequency. The base state variables in the non-dimensional form become:
\begin{equation}
V = W = 0, \quad U = z, \mbox{ and } B = z - \frac{y}{Ri},
\end{equation}
where $\mbox{Ri} = H^2 N^2/u_0^2$ is the Richardson number.
The non-dimensionalized system (\ref{Euler}) once linearized around the base state becomes
\begin{subeqnarray}
\frac{\partial u}{\partial t}+ U\frac{\partial u}{\partial x} + w \frac{\partial U}{\partial z} -v + \mbox{Ri} \frac{\partial p}{\partial x}&=&0,\\
\frac{\partial v}{\partial t}+ U  \frac{\partial v}{\partial x} +u + \mbox{Ri} \frac{\partial p}{\partial y}&=&0, \\
\delta^2 \left(\frac{\partial w}{\partial t} + U \frac{\partial w}{\partial x}\right)  - \mbox{Ri} b + \mbox{Ri} \frac{\partial p}{\partial z} &=&0, \\
\frac{\partial b}{\partial t} + U\frac{\partial b}{\partial x} - \frac{v}{\mbox{Ri}}   +w   &=&0, \\
\frac{\partial u}{\partial x}  + \frac{\partial v}{\partial y}  +\frac{\partial w}{\partial z} &=&0.
\label{Euler_ad}
\end{subeqnarray}
where the second control parameter $\delta = fH / u_0$ is a measure of the aspect ratio $\lambda = H/L$ divided by the Rossby number $\mbox{Ro} = u_0/fL$ as the horizontal length scale is $L=u_0/f$. Note that the deviation from hydrostatic equilibrium is determined by the value $\delta^2 /\mbox{Ri} = f^2/N^2$ and provides a rescaling of the vertical velocity $w$ and the buoyancy perturbation $b$ with respect to the horizontal components $(u,v,p)$ in the case of finite length scales.
This number also corresponds to the squared aspect ratio $\lambda$ when choosing $L = N H/f$ as a horizontal length scale as in \cite{MolemakerM:05}.

\subsection{Stability analysis}

The solution for the stability is sought in term of linear perturbation variables, denoted by the superscript $(\tilde{\;})$ such that:
\begin{equation}
(u,v,w,p,b) =(\tilde{u}(z),\tilde{v}(z),\tilde{w}(z),\tilde{p}(z),\tilde{b}(z))e^{i(\omega t + \alpha x+\beta y)}
\label{modes}
\end{equation}
where $\alpha$ and $\beta$ are zonal and meridional wave numbers, respectively and where $\omega$ is the eigenvalue. Substituting the ansatz (\ref{modes}) in system (\ref{Euler_ad}), the eigenvalue problem consists of a coupled system of ordinary differential equations, expressed in terms of $z$-dependence only for the perturbation variables $(\tilde{u},\tilde{v},\tilde{w},\tilde{p},\tilde{b})$ with
\begin{subeqnarray}
i (\omega + \alpha U)\tilde{u}  - \tilde{v} + \tilde{w} + i \alpha Ri \tilde{p} &=& 0,\\
\tilde{u} + i (\omega + \alpha U) \tilde{v} + i \beta Ri \tilde{p} &=& 0,\\
i \delta^2  (\omega + \alpha U) \tilde{w} + Ri \frac{\mbox{d} \tilde{p}}{\mbox{d} z} - Ri \tilde{b} &=& 0,\\
- \frac{\tilde{v}}{Ri} + \tilde{w} + i (\omega + \alpha U) \tilde{b} &=& 0,\\
i \alpha \tilde{u} + i \beta \tilde{v} + \frac{\mbox{d} \tilde{w}}{\mbox{d} z} &=& 0.
\label{NSNDL}
\end{subeqnarray}
The eigenvalue problem (\ref{NSNDL}) is reduced to three equations for the vertical velocity $\tilde{w}$, buoyancy $\tilde{b}$, and vertical vorticity $\tilde{\eta} \equiv \partial v/\partial x - \partial u/\partial y = i \alpha \tilde{v}-i\beta \tilde{u}$
equations and reads
\begin{subeqnarray}
i \omega \left(\frac{d^2}{dz^2} - k^2 \delta^2\right) \tilde{w} &=& -i \alpha U \left(\frac{d^2}{dz^2} - k^2 \delta^2\right) \tilde{w} - k^2 \mbox{Ri} \tilde{b} + \frac{d \tilde{\eta}}{dz},\\
i \omega \tilde{b} &=& \left(\frac{i \beta}{k^2 \mbox{Ri}}\frac{d}{dz} - 1\right) \tilde{w} - i \alpha U \tilde{b} + \frac{i \alpha}{k^2 \mbox{Ri}} \tilde{\eta},\\
i \omega \tilde{\eta} &=& - \left(i \beta + \frac{d}{dz}\right) \tilde{w} - i \alpha U \tilde{\eta},
\label{vort-eta-b}
\end{subeqnarray}
where $k^2 = \alpha^2+\beta^2$, which in a matrix form become 
\begin{equation}
\omega J \textbf{q} = L\textbf{q} ,
\label{IVP}
\end{equation}
where $\textbf{q} = [\tilde{w}, \tilde{b}, \tilde{\eta}]$, and $J$ and $L$ are matrices given by\\

\[J=
\begin{bmatrix}
    D^2-k^2 \delta^2 & 0 &  0  \\
    0 & I & 0 \\
    0 & 0 & I 
\end{bmatrix}
\]
and
\[L=
\begin{bmatrix}
    -\alpha U (D^2-k^2\delta^2)  & i \mbox{Ri} k^2 &  -iD  \\
    \frac{\beta　D}{Ri k^2} + i& -\alpha U & \frac{\alpha}{Ri k^2} \\
    iD - \beta & 0 & -\alpha U 
\end{bmatrix}
\]
\\
and where $D = \mbox{d}/\mbox{d}z$ and the boundary conditions
\begin{equation}
\tilde{w}\Bigg{|}_{z=0,1}= 0.
\end{equation}
The coupled eigenvalue problem (\ref{vort-eta-b}) is discretised finally using second-order finite differences with $N_z=1000$ equally-spaced grid points along the vertical direction $z$ as in \citet{MolemakerM:05},
giving rise to a matrix-type eigenvalue problem that can later be used to perform transient growth analysis \citep{schmid2012stability}.\\

In the next subsection, we analyze the asymptotic properties of the eigenvalue problem originally formulated in \citet{Stone:71} for the vertical velocity only, equivalent to the system (\ref{vort-eta-b}a-c). In particular, we carry out dispersion relation analyses for this particular system to identify and then separate the resolved from the spurious modes.\\

\subsection{Eigenvalue spectrum and modes selection}

In his original work, \citet{Stone:71} showed that (\ref{NSNDL}) could be combined into a single equation for the vertical velocity which writes
\begin{equation}
\left(1-\phi^2\right)\frac{\mbox{d}^2 w}{\mbox{dz}^2} - 2\left(\frac{\alpha}{\phi} - i\beta\right)\frac{\mbox{d}w}{\mbox{dz}}-\left(\mbox{Ri}(\alpha^2 + \beta^2)-(\alpha^2+\beta^2)\delta^2\phi^2 + \frac{2i\alpha\beta}{\phi}\right)w = 0,
\label{Stone_eq}
\end{equation}
where $\phi(z)=\omega+\alpha U(z)$, and used asymptotic arguments to compute approximations of both geostrophic and ageostrophic modes.  As can be readily seen from the  terms containing $\phi$ in equation (\ref{Stone_eq}), the eigenvalue problem turns out to have multiple critical points that occur at $\phi(z)=[-1,0,1]$. Therefore, one would expect to observe different branches associated with each critical solution in the eigenspectrum. 
This was previously observed by several authors \citep{Nakamura:88,Heifetz:03} who already suggested that certain branches, associated with particular physical phenomena, could be removed.
More precisely, \cite{Heifetz:03} used \citet{Nakamura:88} analysis to identify continuous branches associated with critical layers. \cite{Heifetz:03} chose to reintroduce viscosity to stabilize these eigenfunctions and smooth the jump across the critical points. In their subsequent study, \citet{heifetz2007generalized} show in their figure 1 that there exist a strong non-orthogonality that is likely to deteriorate with increasing the number of discretization points in the matrix used to solve the eigenvalue problem. The interactions between modes that are orthogonal to each other do not contribute to the energy of the system, whereas non-orthogonal modes have additional contribution to the energy gain.
The importance of this orthogonality condition between modes was reinforced by the study of \cite{MolemakerM:05} who also mentioned that a large number of discretization points were necessary to obtain accurate eigenvalues.
In addition, \cite{Heifetz:03} noted that it was necessary to remove modes that do not contribute to transient growth but appear in the solution as non-fully resolved marginally oscillatory modes. These modes, also known as Poincar\'e waves \citep[cf. ][pg. 59-60]{Heifetz:03}, are the inertio-gravity waves in the absence of shear which are pure harmonic waves.\\

In this study, we have chosen not to reintroduce viscosity. Instead, we derive dispersion relations for both inertia-gravity waves and the rotational singular modes, associated with the critical layers at $\phi(z)=[-1,0,1]$, which allows for computing the ranges of their $(\omega_r,\omega_i)$ values. Based on these ranges, it is possible to select the appropriate subspace of eigenmodes from the full eigenspectra to compute the transient growth rates. \\ 

\begin{figure}    
\begin{subfigure}[b]{0.5\textwidth}
                \includegraphics[width=\linewidth]{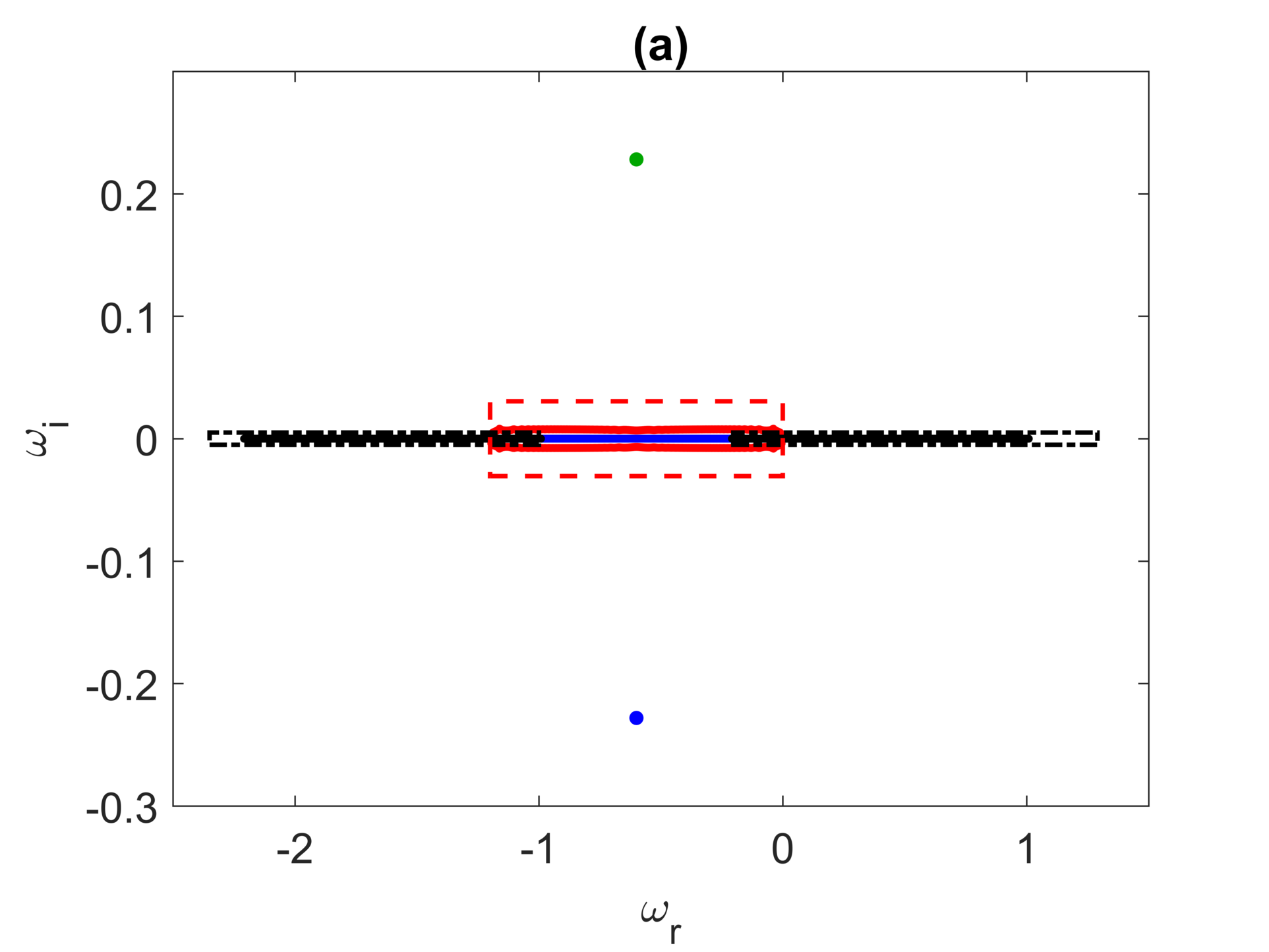}
                \end{subfigure}
 \begin{subfigure}[b]{0.5\textwidth}
                \includegraphics[width=\linewidth]{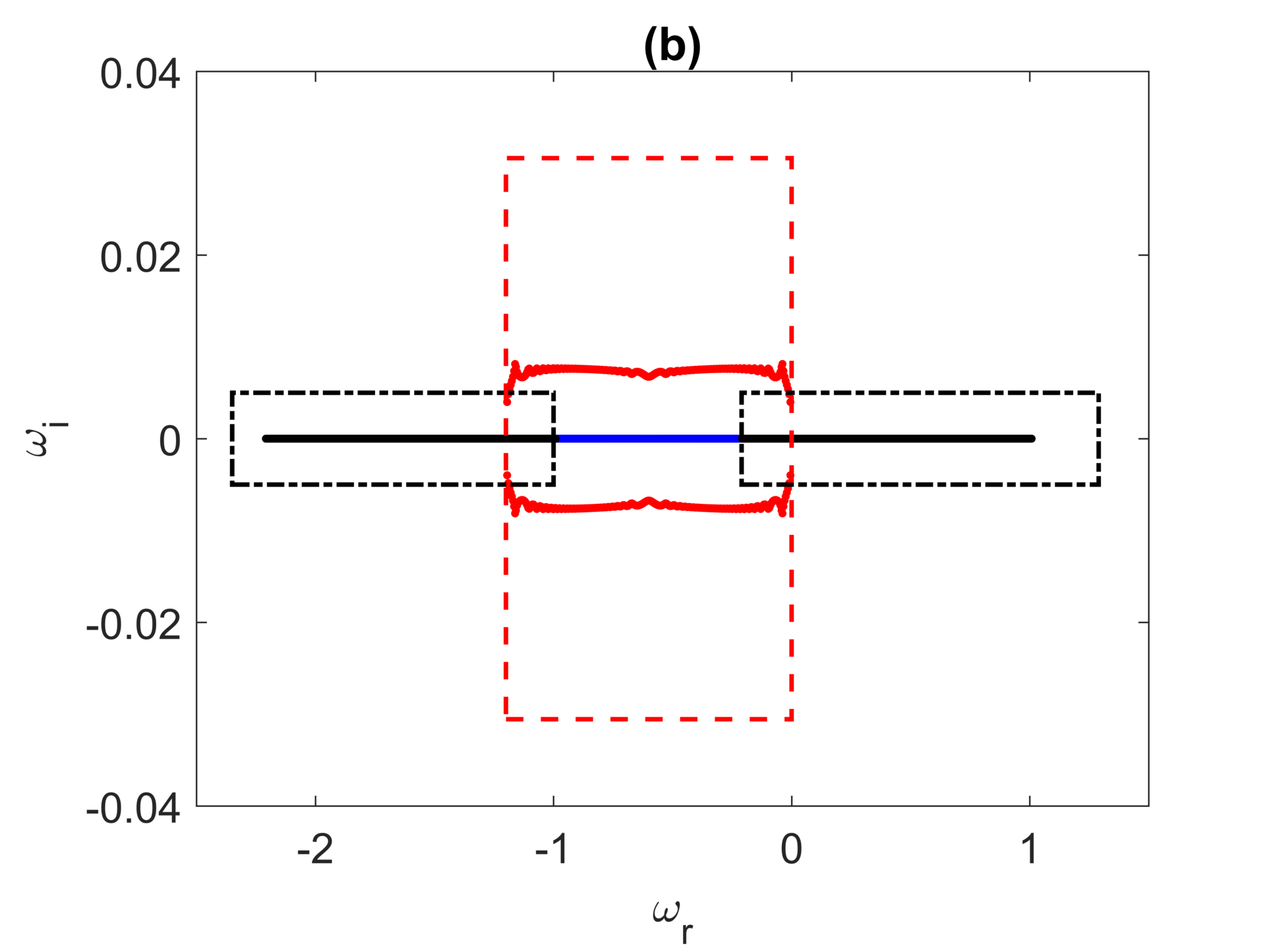}
                \end{subfigure}
                
 \begin{subfigure}[b]{0.5\textwidth}
                \includegraphics[width=\linewidth]{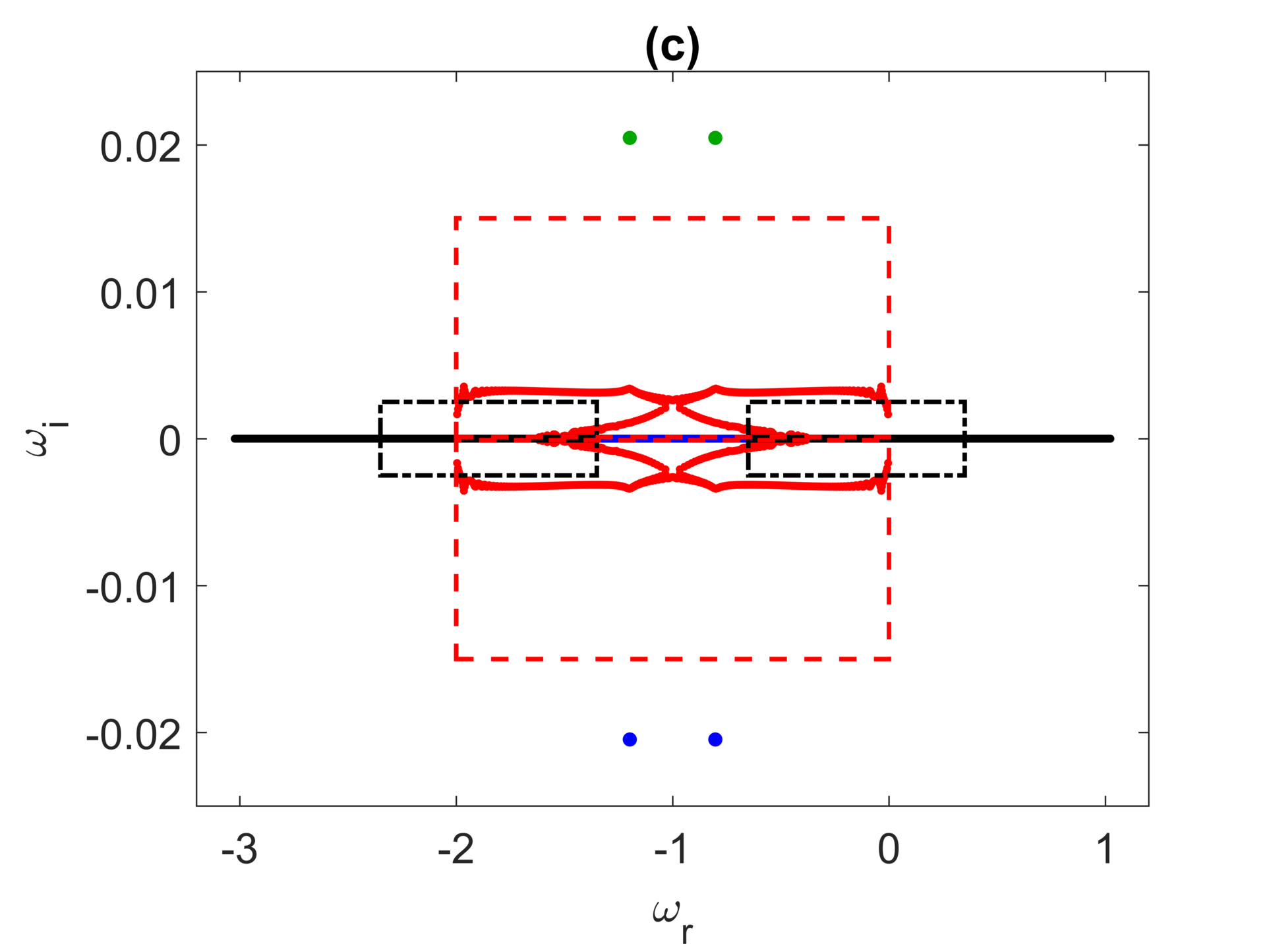}
                \end{subfigure}
 \begin{subfigure}[b]{0.5\textwidth}
                \includegraphics[width=\linewidth]{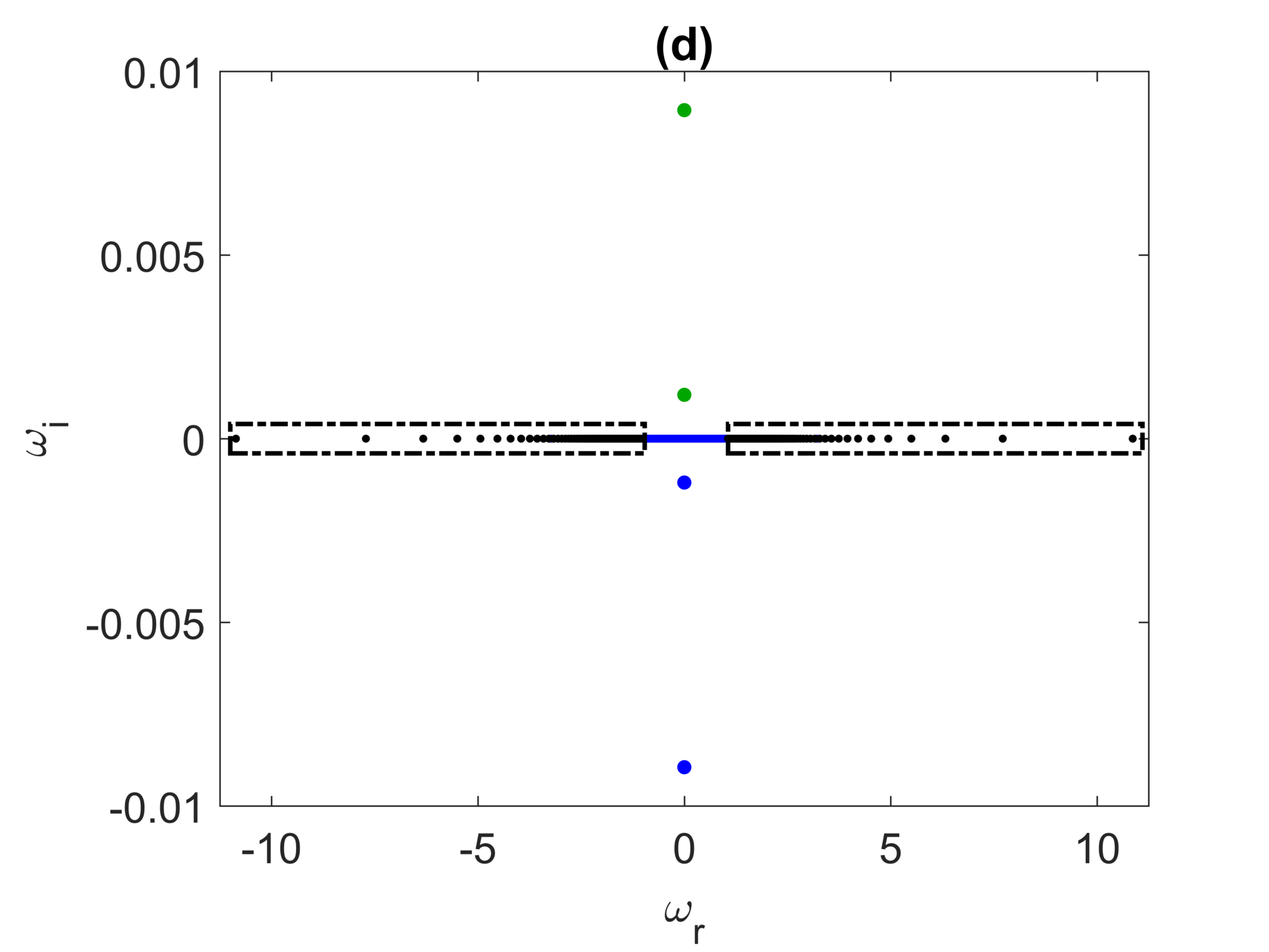}
                \end{subfigure}               
        \caption{Schematic of eigenspectra obtained by solving the eigenvalue problem (\ref{vort-eta-b}) for $\mbox{Ri}=0.92$, $\delta=0$ and (a) geostrophic mode $\alpha=1.2$,  $\beta=0$; (b) close up view of the eigenspectrum from (a) around the real axis; (c) ageostrophic mode $\alpha=2$,  $\beta=0$; (d) symmetric mode $\alpha=0$,  $\beta=12$. Poincar\'e modes (see subsection 2.4) are along the real axis shown in black within a dashed-dotted box, and the singular modes due to critical layers (see subsection 2.5) are shown in red within the dashed box. The modal growth rate, as found in \citet{Stone:71,Stone:70}, is in green, and the remainder of the modes are in blue. In case of symmetric modes, there are no rotational singular modes, but for this $(\alpha,\beta)$ combination, there are two unstable modes ($\omega_r=0, \omega_i>0$). }
        \label{Eigenspectrum}
\end{figure}

Figure \ref{Eigenspectrum} shows eigenspectra for the solution to the system of equations (\ref{vort-eta-b}) for a characteristic case where $\alpha=1.2$, $\beta=0$, $\mbox{Ri}=0.92$, and $\delta=0$, which is a geostrophic mode.
As previously discussed in \citet{Heifetz:03}, the eigenspectrum consists of four separate branches,
which are detailed below, separated into modes that are to be discarded because they are spurious and unresolved or the physical modes that are numerically resolved and used to compute transient growth: 

\begin{itemize}
\item In Figure \ref{Eigenspectrum}(b), two branches are shown by black dots with purely real  frequencies, the ranges for which are calculated in the next subsection.  
These modes are the inertia-gravity waves (i.e. Poincar\'e modes), which are sinusoidal in nature and therefore orthogonal with respect to the other modes. They are associated with critical layers corresponding to $\phi = (-1,1)$, as will be derived in the next subsection.

\item Rotational singular modes are shown by red dots. These modes are associated with critical layers (i.e. $\phi=0$) and have $\omega_r \in [-\alpha, 0]$ and non-zero $\omega_i$. Their frequency and growth rates are within a red dashed box, which will be defined in the following subsection. From \citet{Stone:70}, we know that there is only one unstable mode at these parameter values ($\alpha$, $\beta$, $\mbox{Ri}$, and $\delta$). These additional modes with non-zero growth rates, as will be shown later, are spurious, and arranged in a ring-type structure.

\item The most unstable mode with $\omega_i$ corresponding to the previous calculations by 
\citet{Stone:70,Stone:71} and $\omega_r = -\alpha/2$ is shown in green.

\item The remainder of the spectra is shown in blue and corresponds to physical modes numerically captured by our discretization and whose growth rate is zero as computed by \citet{Stone:70}. \\
\end{itemize}

The geostrophic (fig. \ref{Eigenspectrum}(a,b)), ageostrophic (fig. \ref{Eigenspectrum}(c)), and symmetric (fig. \ref{Eigenspectrum}(d)) modes have eigenspectra with similar anatomy. Note that for symmetric modes, $\alpha = 0$, meaning that the rotational singular modes are not present in the eigenspectrum, and the unstable modes have $\omega_r=0$. For the values of $(\alpha,\beta)$ chosen, there are two unstable modes with $\omega_i>0$, which are discussed in \S \ref{sec:modal_growth}.\\

\begin{figure}
     \includegraphics[width=1.\linewidth]{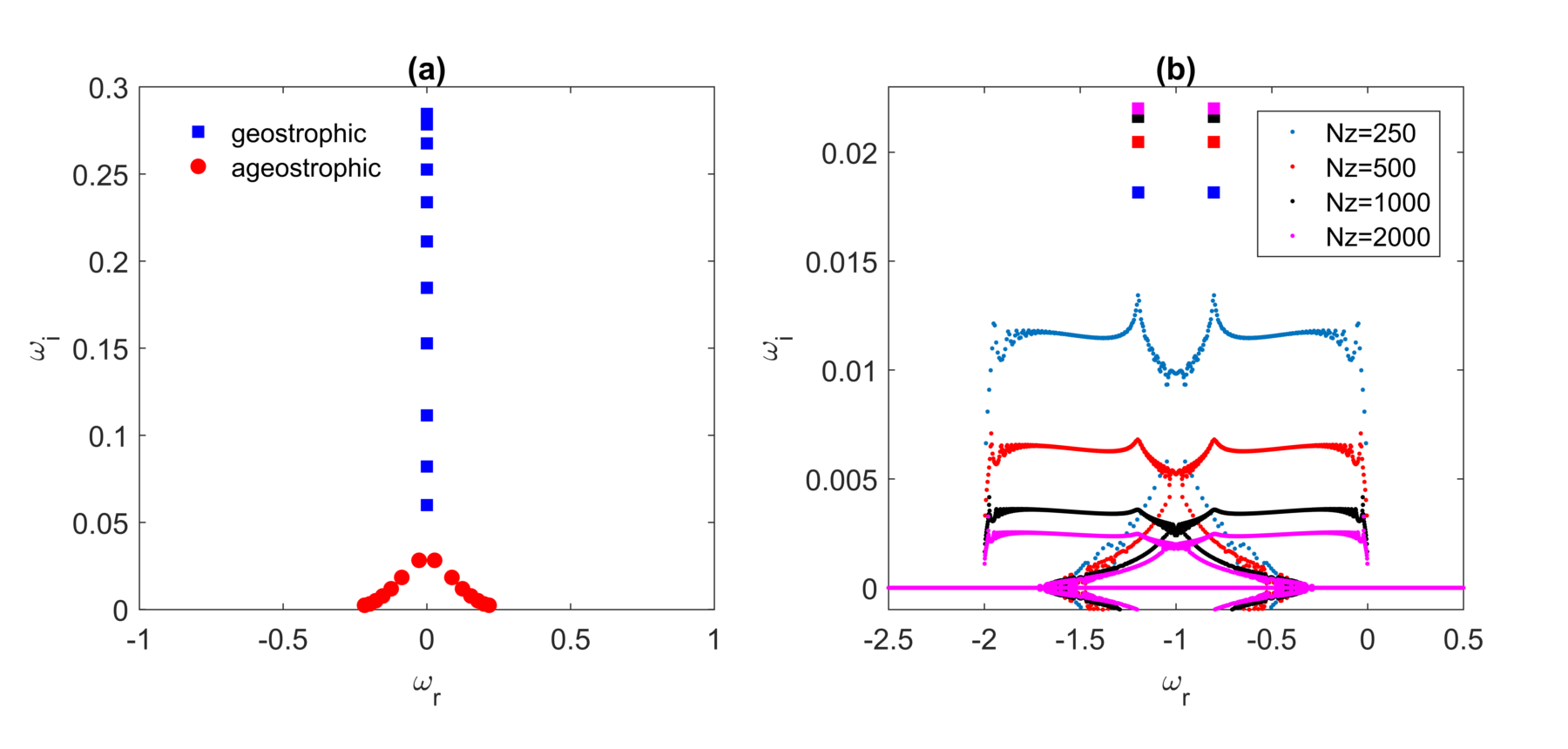}
        \caption{(a) Most unstable eigenvalues rescaled by $\alpha$ for geostrophic (blue squares) and ageostrophic (red circles) modes for $\rm{Ri}=0.5, \delta=0.1, \beta=0$ obtained by varying values of $\alpha$. The region with just one unstable eigenmode corresponds to the geostrophic modes, whereas the two ageostrophic modes with same $\omega_i$ but different $\omega_r$ emerge at larger $\alpha$. (b) Most unstable eigenvalues and the spurious modes for $\rm{Ri}=0.92, \delta=0.1, \alpha=2, \beta=0$ calculated using different number of vertical discretization points $N_z$.}
        \label{EigAna}
\end{figure}

For parameter values corresponding to the ageostrophic modes, there are two modes that have the most unstable growth rates, with $\omega_r$ symmetrically distributed about $-\alpha/2$. These modes result from the two critical layers discussed in \citet{Nakamura:88} in the ageostrophic regime, and are also found in the eigenspectrum for large $\alpha$ by \citet{heifetz2007generalized}. Analogously, \citet{Nakamura:88} finds one critical layer in the geostrophic regime, corresponding to a singular unstable mode shown in figure \ref{Eigenspectrum}(a). The most unstable modes collapse on to a curve when they are rescaled by $\alpha$, such that $(\omega_r-\alpha/2)/\alpha + i \omega_i/\alpha$, which is shown in figure \ref{EigAna}(a) over a range of $\alpha$ values for $\rm{Ri}=0.5$, $\delta=0.1$, $\beta=0$. The structure of the rescaled modes resembles the steering level analysis on \citet{Nakamura:88} with one steering level (singular unstable mode) for the geostrophic modes, and two steering levels (two unstable modes) for the ageostrophic modes. \\

The selection criteria are explained in the following subsections using asymptotic expansions for both the inertial gravity waves and the critical layer modes using dispersion relations. 
The singular behavior of critical layers will translate into spurious eigenvalues and eigenvectors with strong oscillations in the vertical direction. The idea in the following is to identify these spurious modes by finding a range of wave frequencies $\omega$ that represent them in terms of the vertical wavenumber $\gamma$, whose range depends on the number of the vertical discretization points of the domain.

\subsection{Inertia-gravity waves}

Transient growth arises from the non-normality between eigenvectors. Normal modes that do not project onto the initial optimal perturbation may therefore be discarded prior to computing the optimization procedure for transient growth. In particular, modes with a large wavenumber in the vertical direction are likely to
be poorly resolved by the present numerical approach and it is  of advantage to identify and remove these modes from the subspace considered for the transient growth analysis.\\

Inertia-gravity waves, also known as Poincar\'e waves, are normal by nature and can be recovered from Stone's equation. Starting with the dimensional form of (\ref{Stone_eq})
\begin{equation}
\begin{split}
\left(\phi^2 - f^2\right) \frac{\mbox{d}^2 w}{\mbox{dz}^2} -2 \left(\frac{f^2 \alpha (\mbox{d}U/\mbox{dz})}{\phi} - i f \beta \frac{\mbox{d}U}{\mbox{dz}}\right)\frac{\mbox{d} w}{\mbox{dz}} \\
+ \left(-\phi^2 k^2 + N^2 k^2 - \frac{2i \alpha \beta (\mbox{d}U/\mbox{dz})^2}{\phi}\right) w = 0,
\label{Stone_dim}
\end{split}
\end{equation}
we seek shear-independent solutions of (\ref{Stone_dim}) in the near quasi-geostrophic limit (i.e. $\mbox{d}U/\mbox{dz} = \delta = 0$) and the non-dimensional form of (\ref{Stone_dim}) reduces to
the vertical velocity equation
\begin{equation}
\left(1-(\omega + \alpha U)^2\right)\frac{\mbox{d}^2 w}{\mbox{dz}^2}-\mbox{Ri}(\alpha^2 + \beta^2)w = 0,
\end{equation}
which is a modified version of the Poincar\'e wave equation from \citet{Heifetz:03},
with a Doppler shift of the eigenvalue frequencies by $-\alpha U$. These modes correspond to large oscillations in the vicinity of the critical layers associated with $\phi = \omega+\alpha U = [-1,1]$.
The dispersion relation is recovered by setting $\mbox{d}/\mbox{d}z = i\gamma$ and $\mbox{d}^2/\mbox{d}z^2 = -\gamma^2$ and 
\begin{equation}
\omega = -\alpha U \pm \sqrt{1+\frac{\mbox{Ri} (\alpha^2 +\beta^2)}{\gamma^2}}.
\label{intertia-gravity_eq}
\end{equation}
where $\gamma$ is the vertical wave number for the short-wavelength oscillations in the vicinity of the critical layers, which depends on the vertical discretization $N_z$. The modes with this dispersion relation \ref{intertia-gravity_eq} are shown in black in figure \ref{Eigenspectrum} within the dash-dotted boxes.\\

\subsection{Critical layers}

The above dispersion relation allows for discarding the Poincar\'e modes located along the real axis but there remain complex conjugate eigenvalues in the vicinity of the Doppler-shifted frequency, shown in red in figure \ref{Eigenspectrum}. 
These modes were highlighted by \citet{Nakamura:88}, damped using viscosity in \cite{Heifetz:03} and are solutions of (\ref{Stone_eq}) in the near quasi-geostrophic limit.
Following Nakamura's approach (\citeyear{Nakamura:88}), $c = \omega / \alpha$ is the wave speed and $\phi^2 = (\alpha U + \omega)^2 = \alpha^2 (U+c)^2$. Equation (\ref{Stone_dim}) is non-dimensionalized in the neighborhood of the critical altitude where $U+c = U_z H(z-z_c)$:
\begin{equation}
\begin{split}
\left( \rm{Ro}^2\alpha^2(z-z_c)^2-1\right) \frac{\mbox{d}^2w}{\mbox{dz}^2} - 2\left(\frac{1}{z-z_c}+2i\rm{Ro}\beta \right) \frac{\mbox{d}w}{\mbox{dz}} - \\
\left(\alpha^2 k^2 \rm{Ro}^2\delta^2 (z-z_c)^2 +k^2 - \frac{2i\rm{Ro}\beta}{z-z_c}\right) w = 0
\end{split}
\end{equation}
In this limit, the dependence on $\mathcal{O}(\mbox{Ro}^2) \sim \mathcal{O}(z-z_c)^2$ 
which drops as we seek for solutions $z\rightarrow z_c$. In the dimensional form, after dividing through by $-f^2$, equation (\ref{Stone_dim}) becomes
\begin{equation}
\frac{\mbox{d}^2w}{\mbox{dz}^2} + 2 \left(\frac{ \mbox{d}U/\mbox{dz}}{U+c}- \frac{i\beta (\mbox{d}U/\mbox{dz})}{f}\right)\frac{\mbox{d}w}{\mbox{dz}} + \left(-\frac{N^2 k^2}{f^2} - \frac{2 i  \beta (\mbox{d}U/\mbox{dz})^2}{f(U+c)}\right)w = 0.
\label{TG_dim}
\end{equation}
which in non-dimensional form becomes

\begin{equation}
\frac{\mbox{d}^2w}{\mbox{dz}^2} - 2\left(\frac{\alpha}{\omega+\alpha U} - i \beta\right)\frac{\mbox{d}w}{\mbox{dz}} - \left(k^2 \mbox{Ri} + \frac{2i\alpha \beta}{\omega+\alpha U}\right)w = 0.
\label{Nakamura}
\end{equation}
The resulting dispersion relation of the Taylor-Goldstein type equation (\ref{Nakamura}) is found to be the same as the \citet{Eady:49} problem (i.e. the near quasi-geostrophic approximation of (\ref{NSNDL}) from \citet{Nakamura:88}). Thus,
we derive our selection criteria by considering the solution of (\ref{Nakamura}) where again, $\mbox{d}/\mbox{dz} = i\gamma$ and $\mbox{d}^2/\mbox{dz}^2 = -\gamma^2$. This is equivalent to solving the dispersion relation
\begin{equation}
\begin{split}
\omega &= -\frac{\gamma^2 \alpha U + 2i\alpha \gamma + 2\alpha \beta \gamma U + k^2 \mbox{Ri} \alpha U + 2i\alpha \beta}{\gamma^2 + 2\beta \gamma + k^2 \mbox{Ri}} \\
&= -\alpha U - i\frac{2\alpha (\gamma + \beta)}{\gamma^2 + 2\beta \gamma + k^2 \mbox{Ri}}
\label{disp_rel_nakamura}
\end{split}
\end{equation}
The eigenmodes that are solutions to (\ref{Nakamura}) are shown in figure \ref{Eigenspectrum}(a,b) with red circles, and the range for their real and imaginary parts can be obtained from the dispersion relation (\ref{disp_rel_nakamura}) shown by the dashed red boxes.
The range of $\omega_r$ of these spurious modes is always $[-\alpha,0]$ (corresponding to $U=z$ and $z\in[0,1]$), as has been previously noted by \citet{Heifetz:03}, because these modes are associated with the critical layers arising from the singularity $\omega+\alpha U=0$ in (\ref{Stone_eq}). Hence, these modes are spurious, resulting from the numerical discretization of strong oscillations in the vertical direction near critical layers. \\

The range of $\omega_i$ in (\ref{disp_rel_nakamura}) is a function of $\gamma$, which depends on the number of vertical discretization points $N_z$ ($\gamma \geq \pi N_z$ because these oscillations have wavelengths smaller than $\Delta z$). It is worth noting that the growth rates of the spurious modes increase linearly with increasing mesh sizes $\Delta z$.
While we cannot compute the exact values of $\omega_i$ for each of the spurious modes from (\ref{disp_rel_nakamura}) because we do not a priori know the vertical wavenumber $\gamma$ of each oscillation at each critical layer, we can obtain a range of growth rates for these spurious modes that guides our selection of the physical modes, and allows us to draw the dashed red boxes in figure \ref{Eigenspectrum}.
As we increase $N_z$, the number of discretization points in the vertical, the growth rates $\omega_i$ of these spurious modes are dampened but do not converge, unlike those of the physical modes (indicated in green in figure \ref{Eigenspectrum}) that converge with finer discretizations. The decrease in the growth rates of these unphysical  modes is consistent with the dispersion relation (\ref{disp_rel_nakamura}), in which the imaginary part decreases with increasing $\gamma$ (or finer mesh discretization due to larger $N_z$). An example to illustrate this phenomenon is shown in figure \ref{EigAna}(b) for a test case with $\rm{Ri}=0.92$, $\delta=0.1$, $\alpha=2$, $\beta=0$. For these spurious modes, $\omega_i \rightarrow 0$ as $N_z \rightarrow \infty$, so their growth rates (and their large contribution to transient growth) can be eliminated with sufficiently high discretization \citep{MolemakerM:05}, but the computational time for the direct method that solves for all eigenvalues increase with the cubic power of the order of the matrix ($3N_z$ for our reduced system (\ref{vort-eta-b})).
However, utilizing the dispersion relation \ref{disp_rel_nakamura}, we can identify and eliminate the spurious modes at substantially coarser grid resolution without sacrificing the accuracy of the solution.\\

The unphysical modes, such as the ones that we find, are also known to occur when the number of independent variables in the incompressible Navier-Stokes equations is reduced via algebraic constraints \citep{Manning:07}, as is the case in this paper where we obtain the reduced system (\ref{IVP}). 
These modes arise from the numerical approximations, in particular in the vicinity of critical layers, and do not converge as the mesh is refined \citep{Walters:83}. They have previously been found in many hydrodynamic stability problems, including the numerical solutions of the Orr-Sommerfeld equations for viscid shear flows \citep{Gary:70,Orszag:71,Gardner:89}.\\

As a result, the only modes considered for the transient growth analysis are depicted in blue in figure \ref{Eigenspectrum} together with the unstable mode(s) in green. We show in the following that these modes represent an appropriate subspace for inviscid transient growth calculations. The selected spectra avoid spurious eigenvalues and carefully select resolved eigenvalues in a systematic manner.

\section{Transient growth}\label{sec:TG}

The system of equations that, when linearized, forms the eigenvalue problem (\ref{IVP}) \textcolor{red}{and} can be generally expressed as an initial value problem, which in the matrix form writes:
\begin{equation}
\frac{\partial \textbf{q}}{\partial t} = J^{-1} L \textbf{q} = L_1 \textbf{q}.
\end{equation}
Its solution is given by the matrix exponential:
\begin{equation}
\textbf{q} = e^{L_1 t} \textbf{q}_0.
\end{equation}
In the remaining, we seek to maximize the energy gain $G(T)$ defined by 
\begin{equation}
G(T) = \underset{{E_0\neq0}}{\mbox{max}} \left(\frac{E(T)}{E_0} \right) = \underset{{\mathbf{q}_0\neq0}}{\mbox{max}} \left(\frac{\|\textbf{q}(T)\|_E}{\|\textbf{q}_0\|_E} \right)
\end{equation}
for a given time of optimization $T$ over all non-zero initial conditions. The norm $E(t)$ is the integral of the energy at time $t$ and $E_0$ is the energy of the initial condition $\textbf{q}_0$, given by the scalar product $\|\textbf{q}\|_E = \textbf{q}^H M \textbf{q}$. In terms of dimensional variables, $\hat{u},\hat{v},\hat{w},\hat{b}$, it is defined over the computational domain $V$ such that
\begin{equation}
\hat{E} = \|\hat{\textbf{q}}\|_E = \frac{1}{2}\iiint_V \bigg( \hat{u}^2 + \hat{v}^2 + \hat{w}^2 + \frac{\hat{b}^2}{N^2} \,\bigg )\mbox{d}V,
\end{equation}
and is the sum of kinetic and available potential energies \citep{Lorenz:55,Passaggia:17,scotti2019diagnosing}. The non-dimensionalised energy norm is
\begin{equation}
E =  \frac{1}{2}\iiint_V \big ( \tilde{u}^2 + \tilde{v}^2 + \delta^2 \tilde{w}^2 + \mbox{Ri}\, \tilde{b}^2 \big ) \,\mbox{d}V
\end{equation}
and expressing the perturbation velocities $\tilde{u}$ and $\tilde{v}$ in terms of perturbation vertical vorticity $\tilde{\eta} =  i \alpha \tilde{v}-i\beta \tilde{u}$, the non-dimensional energy norm is
\begin{equation}
E  =\frac{1}{2} \int_z \bigg( \frac{1}{k^2}(D^2 + \delta ^2 k^2) \tilde{w}^2 + \frac{1}{k^2}\tilde{\eta}^2 + \mbox{Ri} \, \tilde{b}^2 \bigg )\,\mbox{d}z,
\label{energy}
\end{equation}
with the energy norm matrix\\
\[M=
\begin{bmatrix}
    (\frac{D^2}{k^2} + \delta^2)/2 & 0 &  0  \\
    0 & 1/(2k^2) & 0 \\
    0 & 0 & \mbox{Ri}/2
\end{bmatrix}.
\]
The matrix $M$ is positive definite and has a Cholesky decomposition such that $M = F^H F$ for some matrix $F$ \citep{schmid2012stability}, and the energy norm for some vectors $\textbf{q}_1$ and $\textbf{q}_2$ can be converted to the $\rm{L_2}$-norm such that
\begin{equation}
(\textbf{q}_1,\textbf{q}_2)_E = \textbf{q}_2^H M \textbf{q}_1 = \textbf{q}_2^H F^H F \textbf{q}_1 = (F\textbf{q}_1, F\textbf{q}_2)_2.
\end{equation}
The optimal transient growth can then be expressed as
\begin{equation}
G(t) = \|e^{-it\Lambda}\|_E = \|F e^{-it\Lambda} F^{-1}\|_2,
\end{equation}
where $\Lambda$ is the diagonal matrix containing the eigenvalues $\omega$ obtained by calculating the normal modal growth after discarding the spurious vectors. 
The singular value decomposition of the matrix $F e^{-it\Lambda} F^{-1}$ yields matrices $U$, $\Sigma$, and $V$. The largest singular value corresponds to $G(t)$, whereas the initial conditions associated with the maximum amplification at time $t$ are found in the first column of matrix $V$. The resulting perturbation fields $\textbf{q}(t)$ are contained in the first column of matrix $U$.

\section{Results}
\subsection{Normal Modal Growth}\label{sec:modal_growth}

\begin{figure}
        \begin{subfigure}[b]{0.5\textwidth}
                \includegraphics[width=\linewidth]{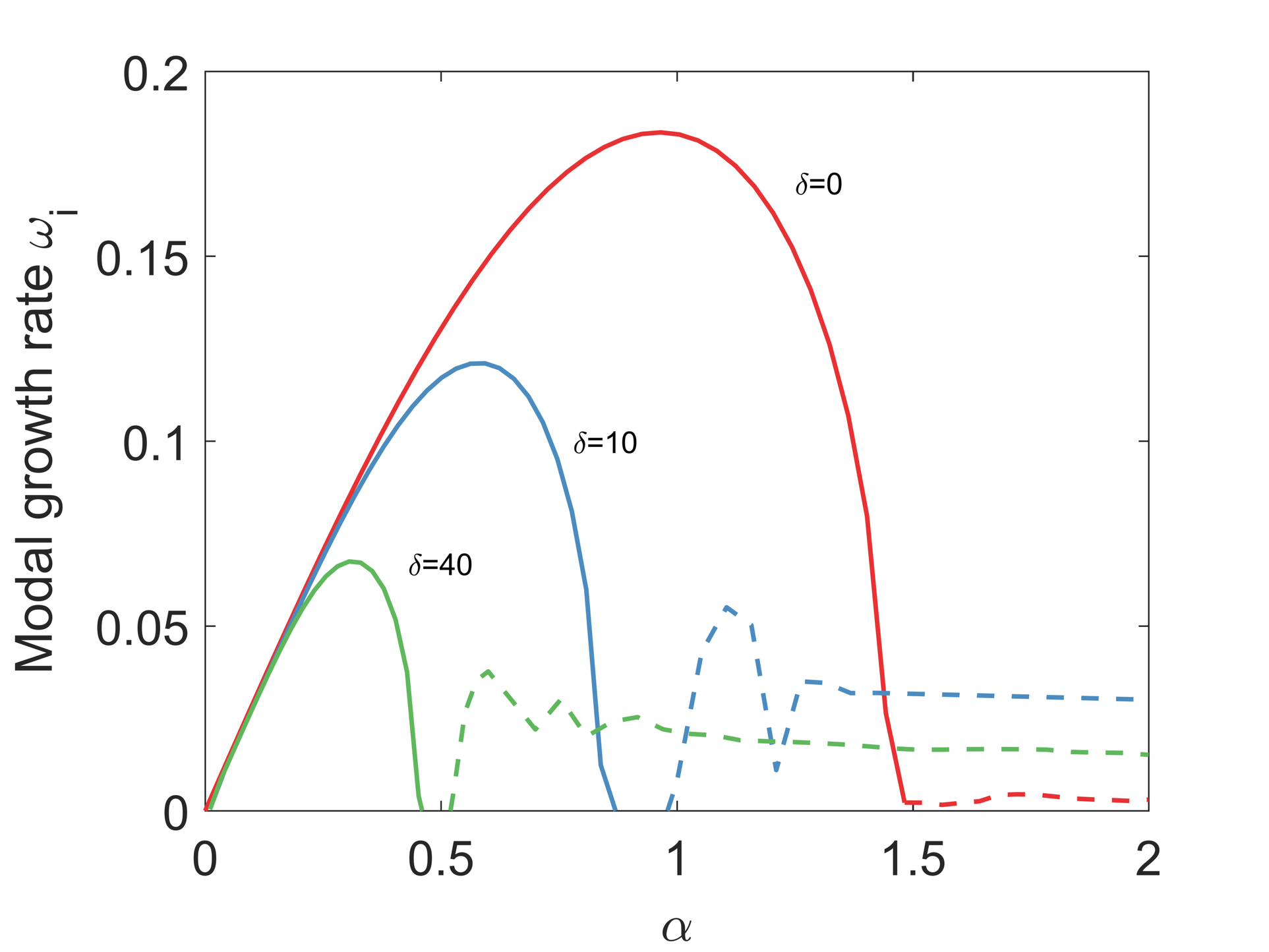}
                \caption{}
                \label{Ri2Stone}
        \end{subfigure}%
        \begin{subfigure}[b]{0.5\textwidth}
                \includegraphics[width=\linewidth]{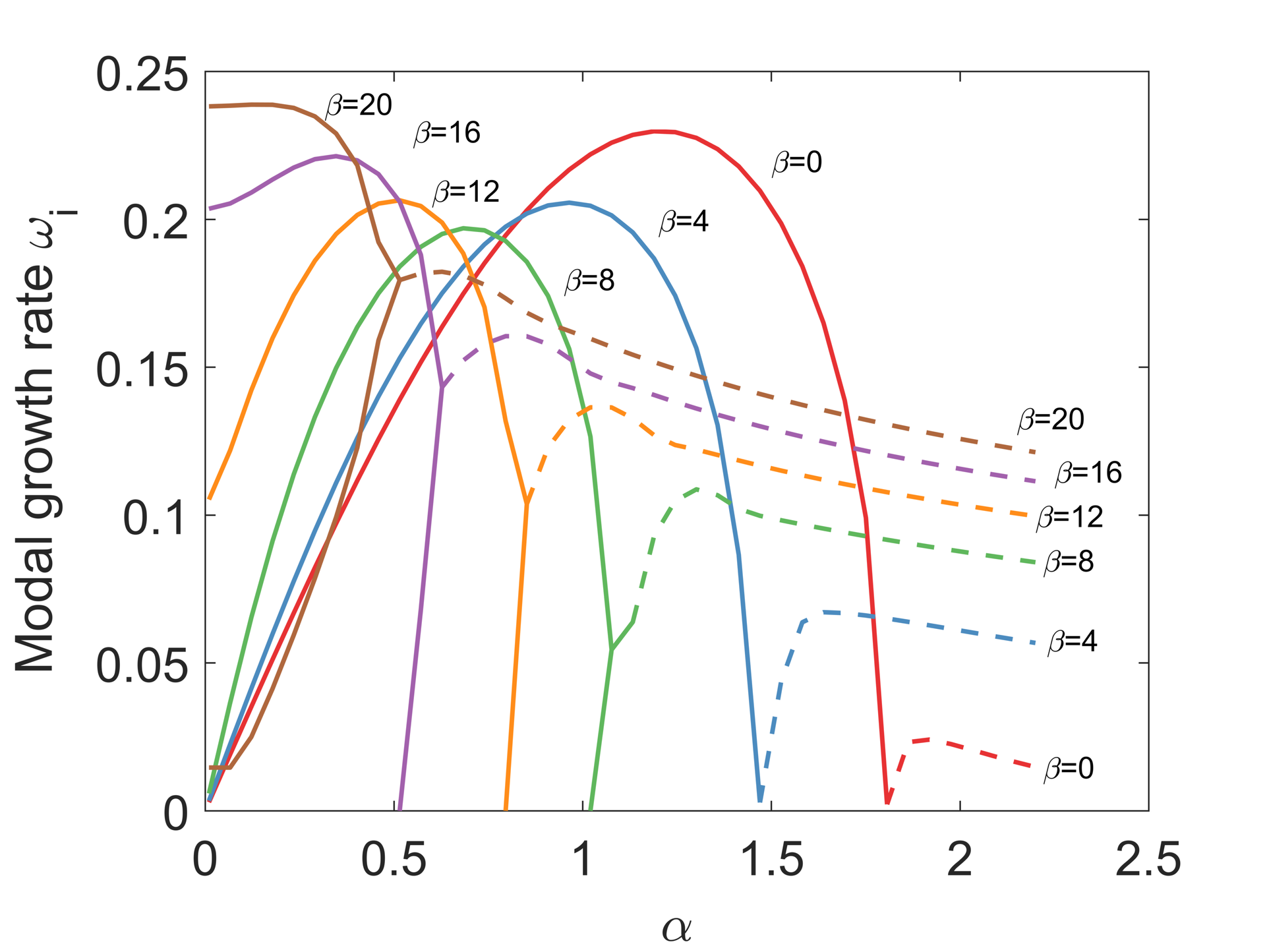}
                \caption{}
                \label{Ri092Stone}
        \end{subfigure}%
        
        \begin{subfigure}[b]{0.5\textwidth}
                \includegraphics[width=\linewidth]{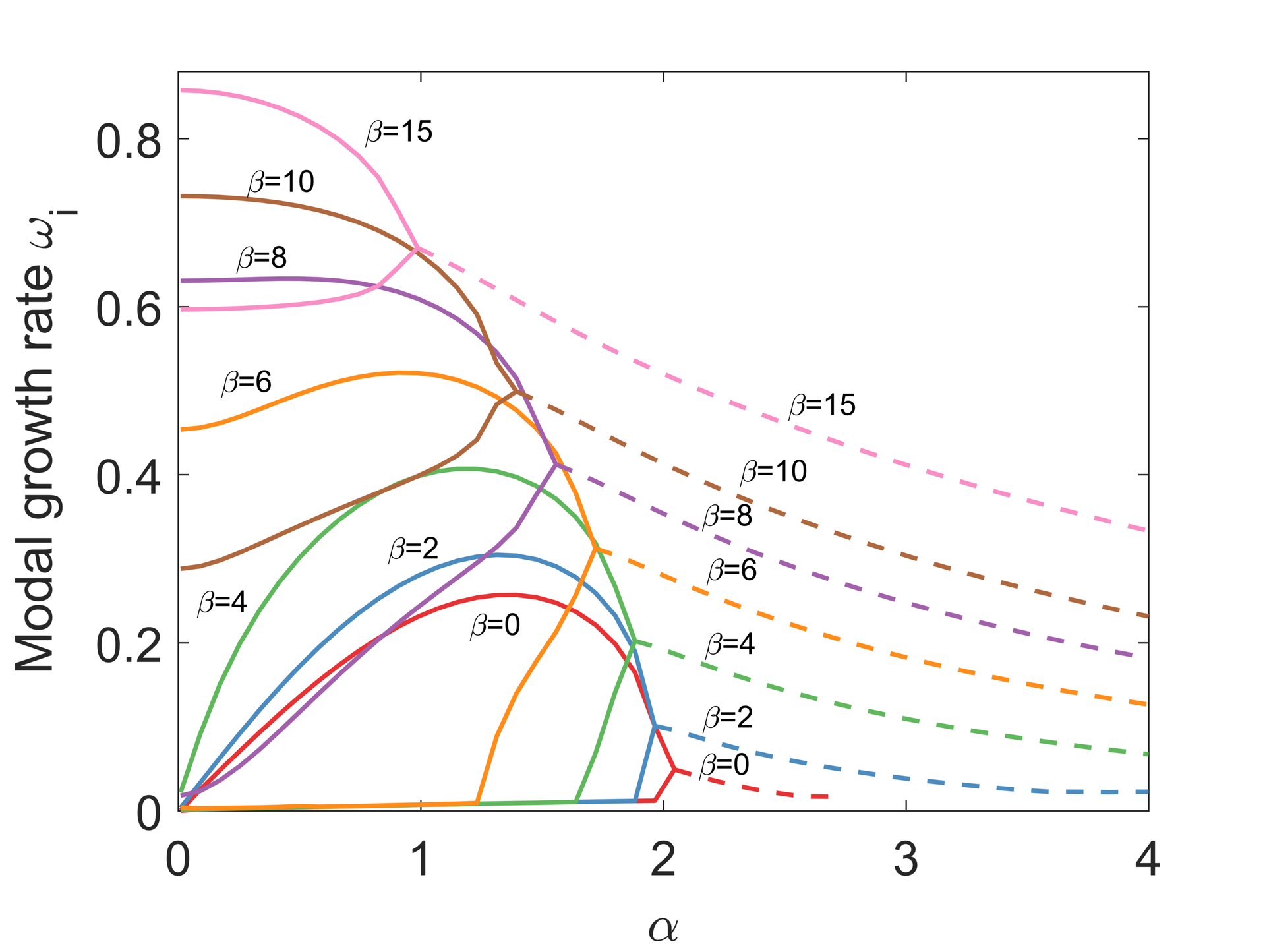}
                \caption{}
                \label{Ri05Stone}
        \end{subfigure}%
        \begin{subfigure}[b]{0.5\textwidth}
                \includegraphics[width=\linewidth]{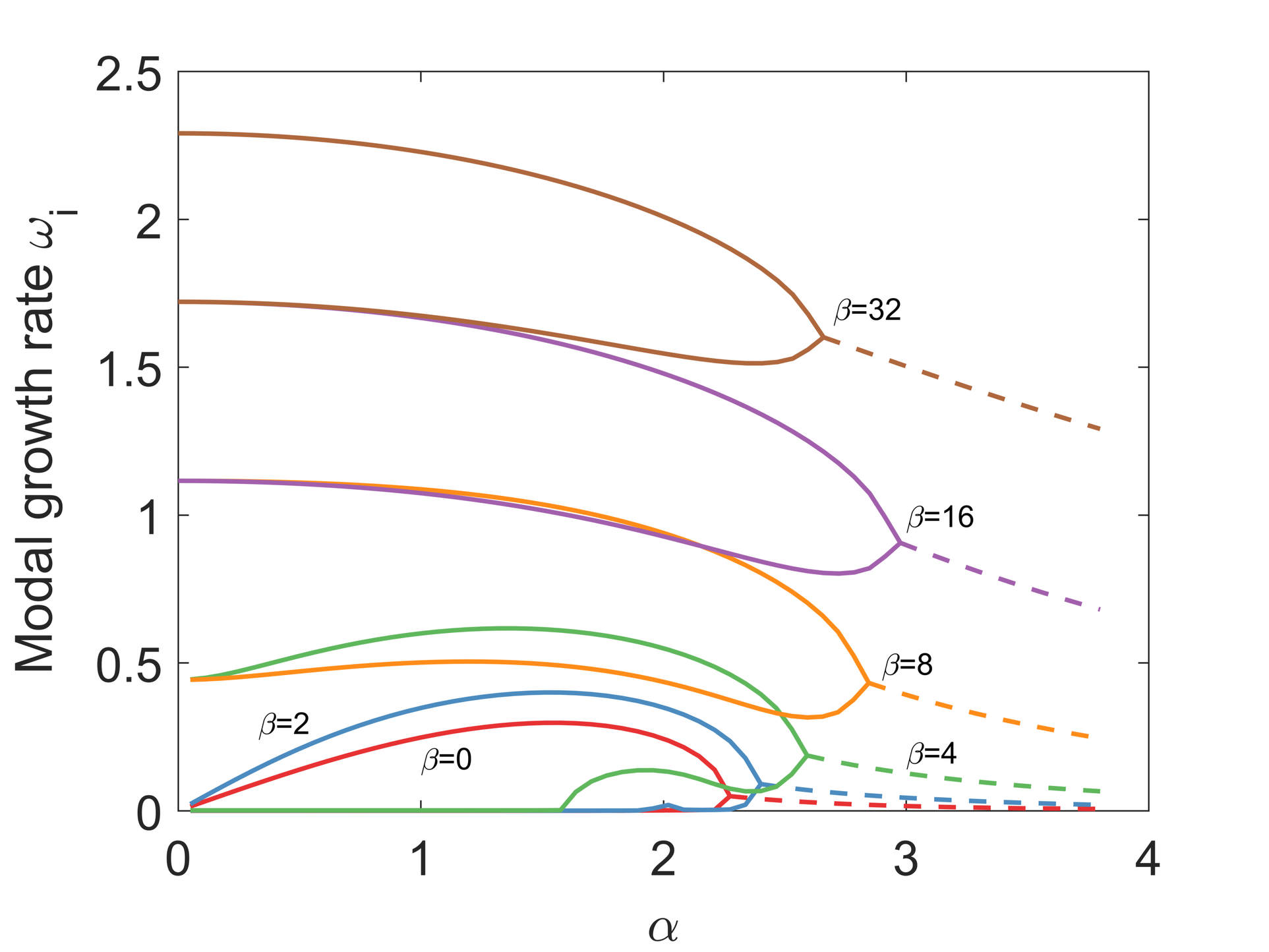}
                \caption{}
                \label{Ri01Stone}
        \end{subfigure}
        \caption{Modal growth rate $\omega_i$ for several values of $\mbox{Ri}$, $\alpha$, and $\beta$. (a) $\mbox{Ri}=2$, $\beta=0$, $\delta=0$ (hydrostatic) and $\delta \neq 0$ (nonhydrostatic) (b) $\mbox{Ri}=0.92$, $\delta=0$, (c) $\mbox{Ri}=0.5$, $\delta=0$, (d) $\mbox{Ri}=0.1$, $\delta=0$. In (c) and (d), both the first and second mode are shown for higher $\beta$ values. Solid lines indicate geostrophic, and dashed lines indicate ageostrophic instabilities.}
        \label{StoneValid}
\end{figure}

We begin with the validation of our approach  and seek to reproduce the theoretical eigenspectrum of \citet{Stone:70,Stone:71} and the numerical solution of \citet{MolemakerM:05} with respect to our solution strategy considering the coupled velocity-vorticity-buoyancy in system (\ref{vort-eta-b}). We employ the finite differences-type scheme used by \citet{MolemakerM:05} who studied the coupled velocity-pressure  system and we calculate the normal mode growth rates $\omega_i$ for different values of $\mbox{Ri}$, $\alpha$ and $\beta$.
Figure \ref{StoneValid}(a) shows $\omega_i$ across $\alpha$ values for $\mbox{Ri}=2$ and $\beta=0$, corresponding to Figure 1 in \citet{Stone:70}, where it is worth noting that the normal mode growth rates are lower under nonhydrostatic conditions ($\delta\neq 0$).
Figures \ref{StoneValid}(b-d) show a sweep over both zonal and meridional wave numbers under hydrostatic conditions for decreasing values of $\mbox{Ri}$, and correspond to figures 3, 4, and 5 from Stone (1970), respectively. 
In $\mbox{Ri}=0.92$ (figure \ref{StoneValid}(b)), we see a transition between $\mbox{Ri} \geq 1$ regime where baroclinic instabilities dominate and the $\mbox{Ri}<1$ regime where 
unstable symmetric modes
become prominent. At lower $\beta$ values, the maximum modal growth rate occurs when $\beta=0$ and decreases with increasing $\beta$ values, and there are no 
unstable symmetric modes
as there is no growth when the zonal wavenumber $\alpha$ is zero. At higher $\beta$ values, this growth rate is non-zero for $\alpha=0$, indicating the presence of 
unstable symmetric modes. At lower $\mbox{Ri}$ values (e.g. $\mbox{Ri}=0.5$, figure \ref{StoneValid}(c)), these instabilities become significant for smaller meridional wave number values.\\

For each given $\beta$ value, there are two sets of instabilities: geostrophic for smaller $\alpha$ values (shown in solid lines) and ageostrophic for larger $\alpha$ values (shown in dashed lines). \citet{Stone:70} discussed the possibility of small ageostrophic growth rates even for $\mbox{Ri}>1$; while \citet{Stone:70} was unable to verify their existence numerically, these smaller growth rates were later calculated by \citet{MolemakerM:05}, who utilized a high-resolution grid, finer than the number of discretization points used in the present study. We also find the ageostrophic modes for $\mbox{Ri}=2$ as shown in figure \ref{StoneValid}(a), which become more significant in the nonhydrostatic regime. \\

For $\mbox{Ri}<1$, there can be more than one unstable eigenmode for certain combinations of $\alpha$ and $\beta$ values. The growth rates for the first two non-zero unstable eigenmodes are plotted in figures \ref{StoneValid}(b-d) for $\mbox{Ri} = 0.92, 0.5$ and $0.1$, respectively. These eigenmodes join together at an intermediate $\alpha$ value and continue as an ageostrophic instability for increasing zonal wave number values $\alpha$. There are 
unstable symmetric modes
present at these values of $\mbox{Ri}$, indicated by the increasing growth rate for $\alpha=0$ and large $\beta$ values. However, in the case of small $\beta$, there still exists the dominance of the geostophic instabilities whose growth rate peaks at $\beta=0$ and decreases with increasing values of $\beta$.
Figure \ref{StoneValid}(d) shows the modal growth rates for $\mbox{Ri}=0.1$. Both first and second eigenmodes are significantly greater than those at lower $\mbox{Ri}$ values for all combinations of $\alpha$ and $\beta$. This regime is dominated by 
unstable symmetric modes
while their growth rates increase with $\beta$ for all values of $\alpha$. \\

\subsection{Non-modal growth}\label{sec:res_TG}

\begin{figure} 
\centering
        \includegraphics[width=1.05\linewidth]{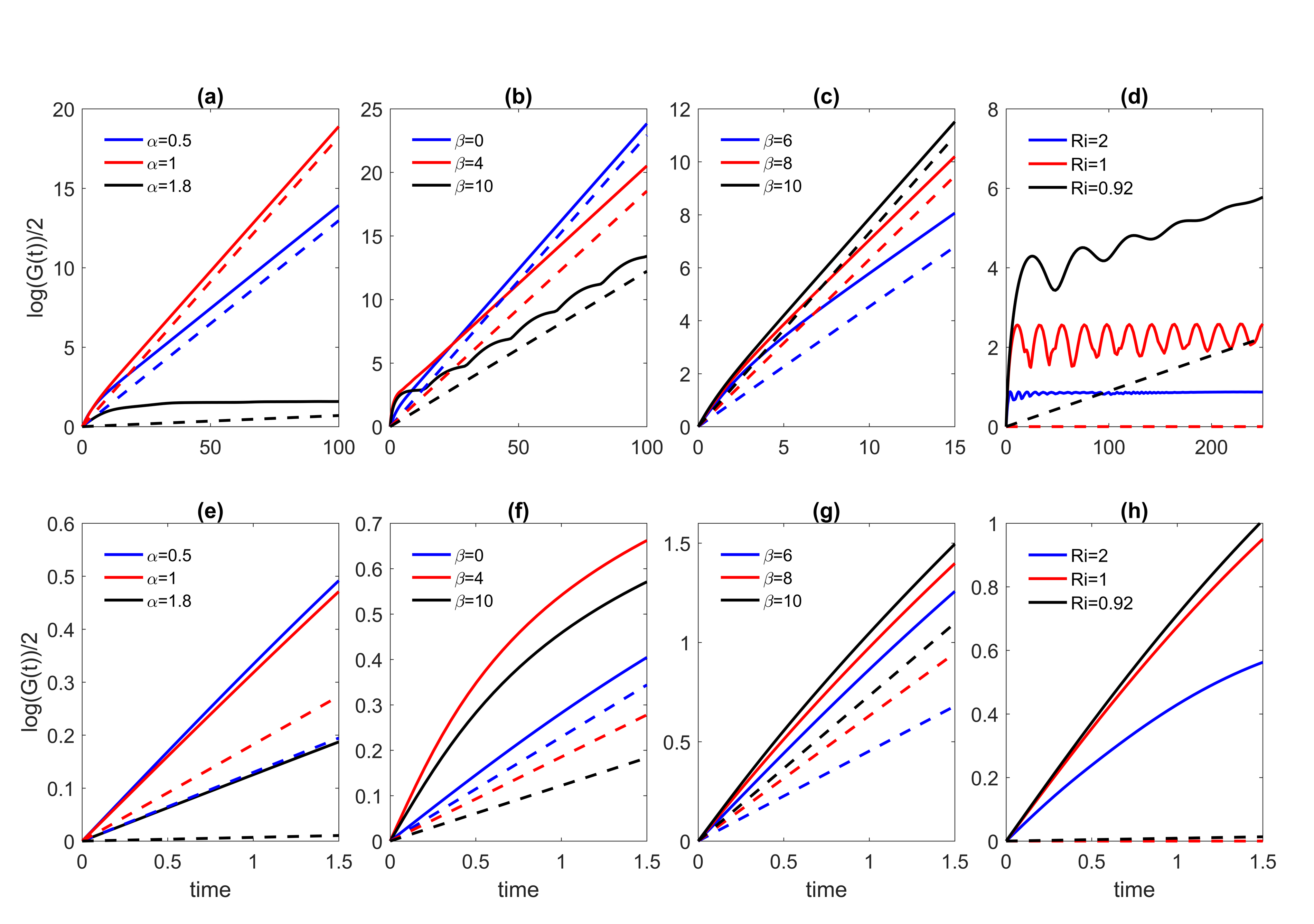}
        \caption{Temporal evolution of energy growth $\mbox{log}(G(t))/2$ (solid) and modal growth $\omega_i t$ (dashed) where the bottom panel figures are zoomed in on the short-time transient growth. Comparison of geostrophic and ageostrophic modes: (a,e) $\mbox{Ri}=2,\, \beta=0$, $\alpha=0.5,1$ geostrophic, $\alpha=1.8$ ageostrophic, (b,f) $\mbox{Ri}=0.92,\alpha=1.2$, $\beta=0,4$ geostrophic, $\beta=10$ ageostrophic. Symmetric modes: (c,g) $\mbox{Ri}=0.5,\alpha=0$, and (d,h) $\alpha=0$ and $\beta=10$ for different $\rm{Ri}$ across the stability boundary predicted by \citet{Stone:70}.} 
        \label{GTimeseries}
\end{figure}

A natural approach to validate the optimal transient growth calculation is to consider the long time asymptotics where for optimization horizons $T\rightarrow\infty$, the associated transient growth rate $\mbox{log}(G(T))/2T$ should asymptotically converge to the most unstable normal-mode growth rate.
Hence we study the long-time dynamics for $T>100$ and observe the optimal energy gain rate approaching the normal growth rate $\omega_i$, calculated in \S\ref{sec:modal_growth}. Figure \ref{GTimeseries} shows the optimal energy growth $\mbox{log}(G(t))/2$ (solid lines) and the normal modal growth $ \omega_i t$ (dashed lines) for different combinations of $\mbox{Ri}$, $\alpha$, and $\beta$. For short optimization times, shown in the zoomed-in bottom panels of figure \ref{GTimeseries}, the energy gain $G(t)$ can exceed the modal energy gain provided by the  modal growth rate quite significantly, indicating the effect of additional non-normal mechanisms, but as $t \rightarrow \infty$, the slope of the optimal growth approaches asymptotically that of the modal growth rate $\omega_i$. 

\begin{figure} 
\begin{center}
     \includegraphics[width=0.8\linewidth]{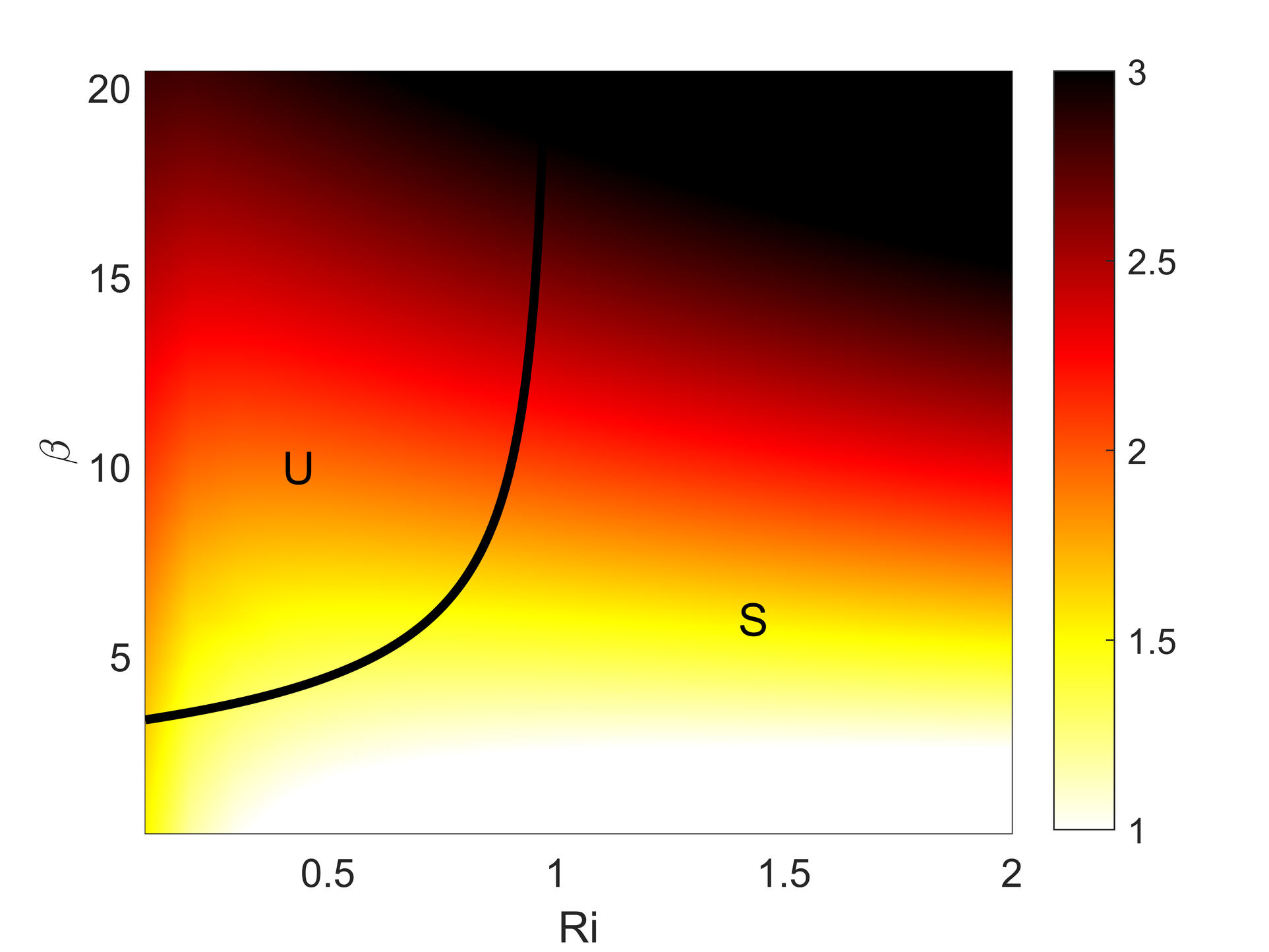}
       \caption{Transient growth rate $\rm{log}G(T=0.5)/2$ for unstable symmetric modes ($\alpha=0$) as function of Ri and $\beta$. The black curve is a theoretical boundary from Stone (1971); no modal growth is predicted for values of Ri and $\beta$ to the right of the line in the region denoted with "S", and nonzero modal growth is found in the region "U".}
     \label{RiBVar}
    \end{center}
\end{figure}

We first compare geostrophic and ageostrophic instabilities of pure baroclinic instabilities ($\beta=0$) across $\alpha$ at $\rm{Ri}=2$ in fig. \ref{GTimeseries}(a,e) and of mixed modes across $\beta$ values with $\alpha=1.2$ at $\rm{Ri}=0.92$ in fig. \ref{GTimeseries}(b,f). Mixed modes refer to the three-dimensional instabilities where $\alpha \neq 0$ and $\beta \neq 0$.
The growth rates of the ageostrophic instabilities are much more amplified by the transient dynamics than those of the  geostrophic modes. At $\mbox{Ri}=2$ (fig. \ref{GTimeseries}(e)), the modal growth rate for the ageostrophic mode ($\alpha=1.8$) is 18-25 times smaller than the rates of geostrophic modes ($\alpha=0.5,1$), but its short time optimal energy growth is only one-third of that of the geostrophic modes. For the mixed modes at $\mbox{Ri} = 0.92$ (fig. \ref{GTimeseries}(f)), the ageostrophic mode ($\beta=10$) has optimal energy gain comparable to that of the mixed geostrophic mode ($\beta=4$) and greater than that of the pure geostrophic mode ($\beta=0$)  at short time, despite having half the modal growth rate. \\

The temporal evolution of the energy growth for the 
unstable symmetric modes
($\alpha=0$) are shown in fig. \ref{GTimeseries}(c,g) at $\rm{Ri}=0.5$. The highest rate of the energy growth rate amplification by transient dynamics occurs at low $\beta$ value: short-term optimal energy gains for symmetric modes with increasing $\beta = [6,8,10]$ are almost equal even though the modal growth rates vary significantly at $[0.45,0.63,0.73]$ for each of the $\beta$ respectively.
Figure \ref{GTimeseries}(e,h) shows 
unstable symmetric modes
$\beta=10$ for different $\rm{Ri}$ across the stability boundary (see figure \ref{RiBVar}) from \citet{Stone:70}. While the linear stability analysis predicts no energy growth at $\rm{Ri}>1$, we observe considerable non-zero transient energy growth at $\rm{Ri}=1,2$ for several inertial periods.
Transient growth rates at $T=0.5$ for 
unstable symmetric modes
over a range $\mbox{Ri}=[0.1,2]$ and $\beta=[0.1,20]$ are shown in figure \ref{RiBVar}. This optimization target time is chosen here as it represents  half an inertial period $f^{-1}$, which is within the range of the timescale for which the submesoscale instabilities occur (compared with the mesoscale instabilities that occur at timescales greater than the inertial period $f^{-1}$) \citep{Boccaletti:07}. The black curve is a theoretical boundary calculated from equation (4.8) in \citet{Stone:71}, which prescribes values of $\beta$ as a function of $\mbox{Ri}$ where modal growth rates decrease down to zero. For values of $\mbox{Ri}$ and $\beta$ to the right of this curve (region marked by "S"), modal growth rates for 
unstable symmetric modes
are zero, but as figure \ref{RiBVar} shows, transient growth rates can be significant outside of this boundary, even for $\mbox{Ri}>1$.\\

\begin{figure}    
     \includegraphics[width=1.\linewidth]{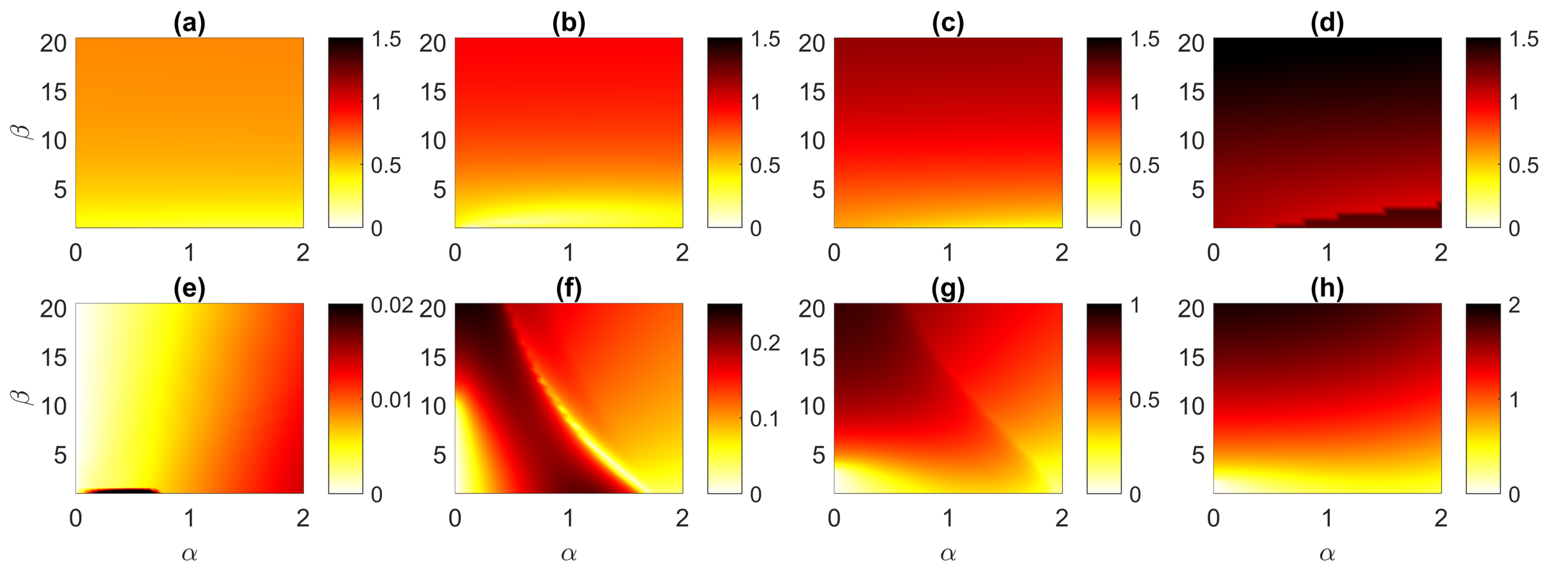}
        \caption{(Top) transient and (bottom)  normal growth rates as functions of $\alpha=[0,2]$ and $\beta=[0.3,20]$. (a),(e) $\mbox{Ri}=2$, (b),(f) $\mbox{Ri}=0.92$, (c),(g) $\mbox{Ri}=0.5$, and (d),(h) $\mbox{Ri}=0.1$. Transient growth rates $\rm{log}(G(T))/2T$ are found at $T=0.5$. }
        \label{TGMGParam}
\end{figure}

We conduct parameter sweeps over $\alpha=[0,2]$ and $\beta=[0.3,20]$ for transient (at $T=0.5$) and normal growth rates shown in the top and bottom panels of figure \ref{TGMGParam}, respectively. The growth rates are computed at $\mbox{Ri} = 2$ (a,e), $0.92$ (b,f), $0.5$ (c,g), and $0.1$ (d,h) to compare different regimes.  As $\mbox{Ri}$ decreases, both transient growth and normal growth rates increase, but the amplification of the normal growth rate  by the transient dynamics decreases. In general, we find that transient growth rates at a given $\mbox{Ri}$ are mostly independent of $\alpha$ and increase with $\beta$, while the normal growth rates are more complex functions of both $\alpha$ and $\beta$. Transient growth especially exceeds modal growth at high Richardson numbers for modes with moderate to large meridional wavenumbers $\alpha$, including the three-dimensional modes where $\beta \neq 0$. In contrast, pure geostrophic modes (small $\alpha$, $\beta=0$) do not seem to gain as much from the transient growth. The transient growth rates of 
unstable symmetric modes
are much larger (up to several orders of magnitude) than the modal growth rates at higher $\rm{Ri}$ and/or at lower $\beta$. When $\rm{Ri}$ is small (fig. \ref{TGMGParam}(c,d,g,h)), 
unstable symmetric modes
have substantial modal growth rates, and the transient growth rates are only larger by about a factor two. \\

These results are in agreement with the study of \citet{heifetz2007generalized} who considered few cases of horizontally isotropic perturbations such as ($\alpha=\beta=[0.1,1,10]$) at $\rm{Ri}=1$. In the case $\alpha=\beta=1$, they report a maximum instantaneous growth rate $\mbox{log}(G)\approx0.5$ at $\rm{Ri}=1$ which is very close to the value of $0.49$ that we report in figure \ref{TGMGParam}(b) at $T=0.5$. We find transient growth rates of $0.26$ and $2.0$ for $\alpha=\beta=0.1$ and $\alpha=\beta=10$, respectively, which also agree with values in \citet{heifetz2007generalized} ($0.26$ and $2.2$, respectively).
Note that \citet{Heifetz:03,heifetz2007generalized,heifetz2008non} report the instantaneous transient growth rates (i.e. at $T=0$), whereas we compute optimal growth rates at $T=0.5$, which is lower than the instantaneous growth rate.

\subsection{Non-hydrostatic effects on transient growth}

\begin{figure}    
     \includegraphics[width=1.\linewidth]{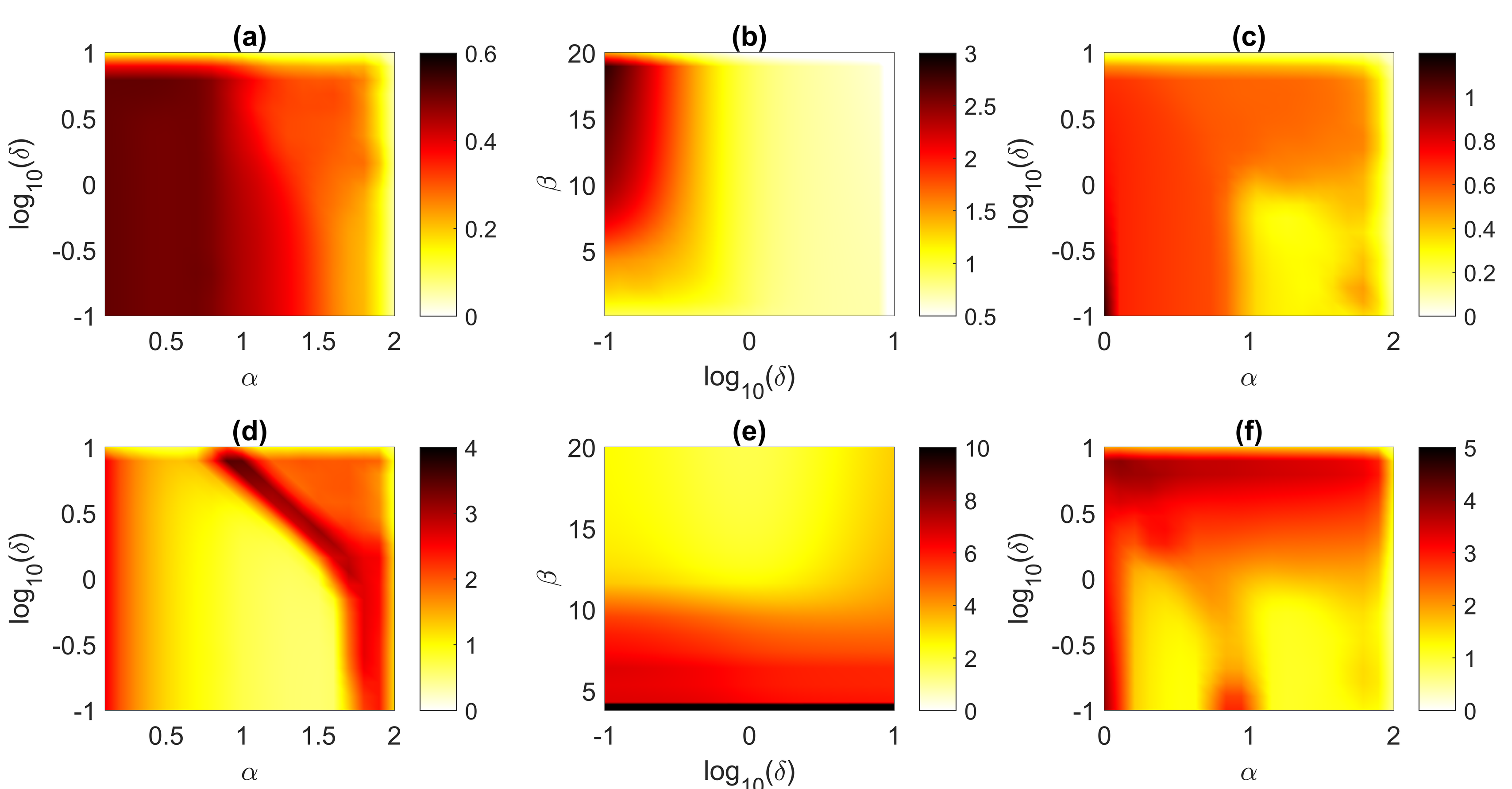}
        \caption{(Top) transient growth rates $\mbox{log}(G(T))/T$ and (bottom) log plot of $\mbox{log}(G(T))/\omega_i T)$ at $T=0.5$ as functions of $\alpha=[0,2]$, $\beta=[0,20]$ and $\delta=[0.1,10]$ for $\mbox{Ri}=0.92$. (a),(d) baroclinic instabilities with $\beta=0$, (b),(e) 
        unstable symmetric modes
        with $\alpha=0$, (c),(f) mixed modes with $\beta=10$. In (e), $\beta<4$ are truncated as the modal growth rates approach zero, so the amplification by the transient growth approaches infinity.}
        \label{NonHydrGR}
\end{figure}

The non-hydrostatic effects have been previously overlooked, mainly because the coupled system (\ref{NSNDL}) in the $\delta\rightarrow0$ limit allows for reducing the number of variables and efficiently solve the associated eigenvalue and generalized stability problems (cf. \citet{Heifetz:03,heifetz2007generalized,heifetz2008non}).
In this section, we study the nonhydrostatic effects by varying the parameter $\delta = (H/L)/\mbox{Ro} = fH/u_0$, which for a fixed $\mbox{Ri}$ is equivalent to varying the aspect ratio $H/L$. In an ocean mixed layer of $\mathcal{O}(100m)$ at mid-latitudes $\delta$ values can reach up to $0.1$, which corresponds to the ocean values reported in \citet{Boccaletti:07} off the coast of California and in the Gulf Stream \citep{callies2015seasonality}.\\

However, the dynamics at $\delta \sim 1$ or greater may be significant, for instance 
in the eye wall region in the case of tornadoes \citep{nolan2017tornado} and hurricanes \citep{worsnop2017gusts}. 
The average aspect ratios of the hurricane boundary layer rolls reported in observations for the Atlantic hurricanes \citep{foster2013signature} and Pacific typhoons \citep{ellis2010helical} fall around $H/L \simeq 0.5$ but can exceed unity, and these rolls can have significant impact on the vertical momentum flux in hurricanes \citep{morrison2005observational}.\\

Using asymptotic analysis, \citet{Stone:71} showed that for baroclinic instabilities, modal growth rates are reduced as $\delta$ increases (see figure \ref{StoneValid}(a)). 
We, therefore, analyzed the effect of non-zero values of $\delta$ on the transient growth dynamics, which generalizes the findings of \citet{Heifetz:03,heifetz2007generalized,heifetz2008non}. While for the geophysical applications, $\delta \simeq 1$ may be the upper bound, we generalize our findings for $\delta=[0,10]$ to fully investigate the nonhydrostatic effects.
Top panels in figure \ref{NonHydrGR} shows how transient growth rates $\mbox{log}(G(T))/2$ found at $T=0.5$ are affected by $\delta$, and bottom panels show the amplification of modal growth rates by optimal linear dynamics for selected parameter spaces at $\mbox{Ri}=0.92$. Figures \ref{NonHydrGR}(a,d) focus on baroclinic instabilities with $\alpha = [0,2]$, figures \ref{NonHydrGR}(b,e) on 
unstable symmetric modes
with $\beta = [0,20]$, and figures \ref{NonHydrGR}(c,f) explore mixed modes with $\beta =10$ and $\alpha = [0,2]$. The transient growth rates are mostly independent of $\delta$ and are larger at lower $\alpha$ values. The structure of $\mbox{log}(G(T))/2\omega_i$ is influenced by the dependence of modal growth rates on $\alpha$ and $\delta$. 
Transient growth rates do not change significantly with $\beta$ or $\delta$ values, except in the small $\delta$ regime in which $\mbox{log}(G(T))/2$ increases with $\beta$. At low $\beta$ values for all $\delta$, $\omega_i \simeq 0$, whereas transient growth rates are nonzero. The amplification of modal growth rates by optimal perturbation decreases with $\beta$ with less significant dependence on $\delta$.\\


\subsection{Energy budget of transient and modal growth}\label{sec:Budget}

We further investigate the effect of transient growth on the energy transfers with an emphasis on the partition between shear production and buoyancy fluxes. One of the important aspects of transient growth is to provide alternative mechanisms for vertical transport or the vertical restratification. In particular we quantify the ratio of horizontal buoyancy flux to the production of kinetic energy. While the first redistributes potential energy in the horizontal direction, the second relates to the amount of vertical transport.\\

From (\ref{energy}), the rate of change of energy in non-dimensional form is
\begin{equation}
\frac{\partial {E}}{\partial t} =  \frac{1}{2} \frac{\partial}{\partial t}\int_V \bigg ( 2 {u} \frac{\partial {u}}{\partial t} + 2 {v} \frac{\partial {v}}{\partial t} + 2 \delta^2 {w} \frac{\partial {w}}{\partial t} + 2 \mbox{Ri} {b} \frac{\partial {b}}{\partial t} \bigg )dV.
\end{equation}
Taking $\zeta = e^{i(\alpha x +\beta y +\omega t )}$, we have ${q} = \tilde{q}(z)\zeta$ and ${q}_t = i\omega \tilde{q}(z) \zeta$.

The equation for the rate of change of energy can be re-written as:
\begin{equation}
\frac{\partial {E}}{\partial t} = i\omega\int_x \int_y \zeta^2 \int_z \big ( \tilde{u}^2 + \tilde{v}^2 + \delta^2 \tilde{w}^2 + \mbox{Ri} \tilde{b}^2\big ) \, dz dy dx.
\end{equation}
Multiplying the non-dimensional linearized governing equations (\ref{NSNDL})(a-d) by $\tilde{u}, \tilde{v}, \tilde{w}, \mbox{ and } \tilde{b}$ respectively, we get an expression for the energy rate of change
\begin{equation}
\frac{\partial {E}}{\partial t} = i\omega\int_x \int_y \zeta^2 \int_z \big (- \tilde{u} \tilde{w} + \tilde{v} \tilde{b} -i\alpha U {E} \big )\, dz dy dx,
\label{energy_norm}
\end{equation}
where the three terms on the right are shear production, meridional buoyancy flux, and horizontal energy transport. Here, the terms are non-dimensional and the background shear $dU/dz=1$ in the shear production term $-uw \,\, dU/dz$. 
It should be noted that the last term does not play a role in the energy balance equation and only characterizes the zonal advection of energy. 
This analysis is somewhat different than the analysis of \citet{Stamper:17} where they only characterize the kinetic energy budget. Instead we consider the total energy budget where the vertical buoyancy flux cancels  out. \\ 

\begin{figure}    
     \includegraphics[width=1.\linewidth]{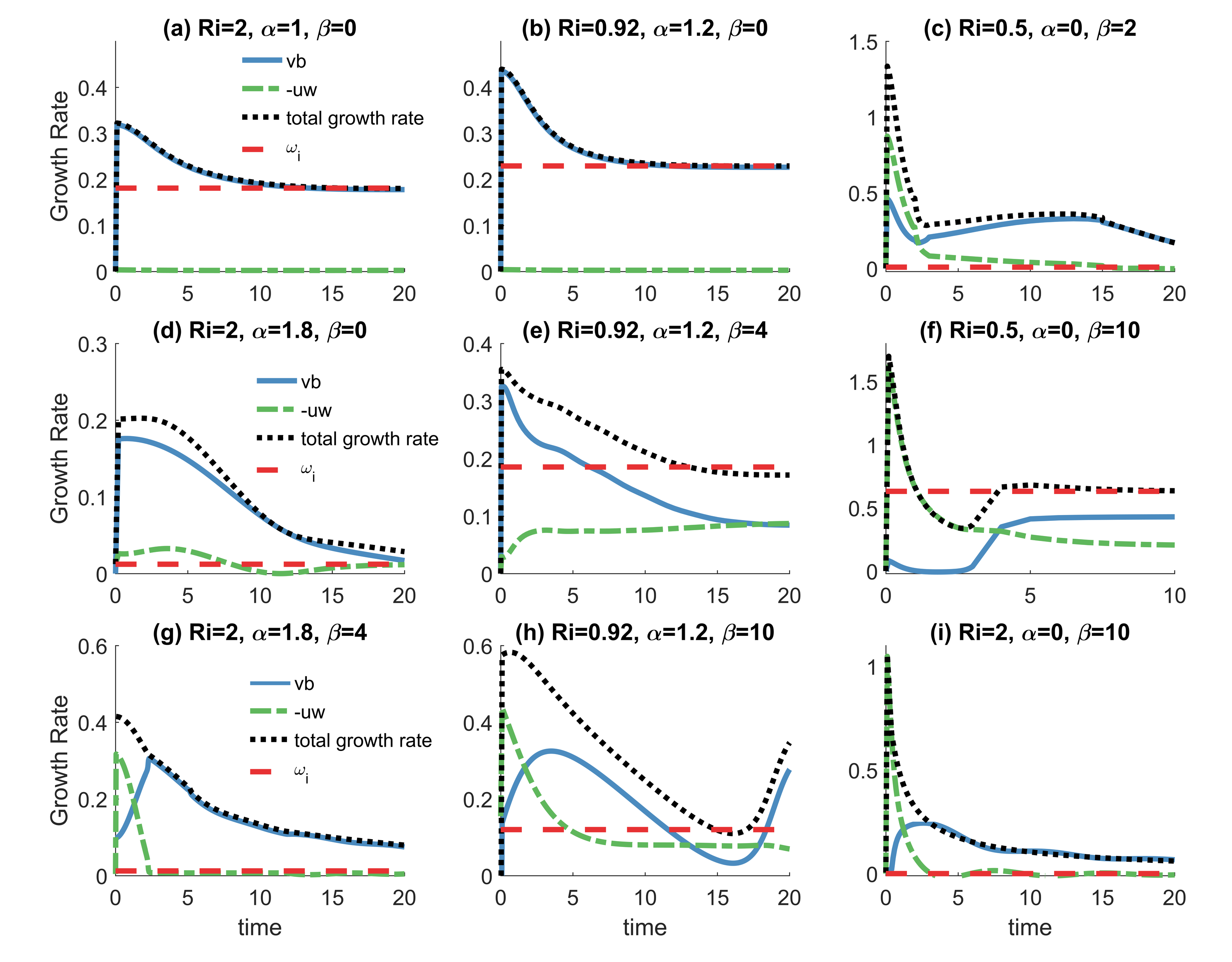}
        \caption{Optimal energy growth rate at optimization time $t$ (meridional buoyancy flux: blue solid; shear production: green dash-dotted; total: black dotted lines) and modal growth rate $\omega_i$ (red dashed line). (Left) $\rm{Ri}=2$ (a) geostrophic mode ($\alpha=1, \beta=0$), (d,g) ageostrophic modes ($\alpha=1.8, \beta=0,4$); (middle) $\rm{Ri}=2, \alpha=1.2$ (b,e) geostrophic modes ($\beta=0,4$), (h) ageostrophic mode ($\beta=10$); (right) symmetric modes ($\alpha=0$) (c) $\rm{Ri}=0.5, \beta=2$, (f) $\rm{Ri}=0.5, \beta=10$, (i) $\rm{Ri}=2, \beta=10$. All values are computed for $\delta=0.1$.}
        \label{GRTimeseries}
\end{figure}

Figure \ref{GRTimeseries} shows the time series of instantaneous growth rate split between the meridional buoyancy flux and shear production along with the modal growth rate for selected cases presented in figure (\ref{GTimeseries}). The left panel plots show (a) geostrophic ($[\alpha,\beta]=[1,0]$) and (d,g) ageostrophic ($[\alpha,\beta]=[1.8,0]$, $[\alpha,\beta]=[1.8,4]$) modes for $\mbox{Ri}=2$. The middle panel shows (b,e) geostrophic ($[\alpha,\beta]=[1.2,0]$, $[\alpha,\beta]=[1.2,4]$) and (h) ageostrophic ($[\alpha,\beta]=[1.2,10]$) modes for $\mbox{Ri}=0.92$. The right panel plots show symmetrically unstable modes ($\alpha=0$) for (c,f) $\mbox{Ri}=0.5$ with $\beta$ values: (c) $\beta=2$ and (f) $\beta=10$, and (i) $\rm{Ri}=2, \beta=0.$
As $t \rightarrow \infty$, the growth rate approaches $\omega_i$, validating our calculations. \\

For geostrophic modes at $\mbox{Ri}>2$ (figure \ref{GRTimeseries}(a)) and $\mbox{Ri}<1$ (figure \ref{GRTimeseries}(b)), the meridional buoyancy flux dominates over the the shear production, which is close to zero, consistent with results presented in the middle panels of figure 7 in \citet{heifetz2007generalized} at $\mbox{Ri}=1,\alpha=\beta=1$. In contrast, in the case of ageostrophic modes, for which energy growth is much more amplified by the optimal linear dynamics than for geostrophic modes, the proportion of energy gain from shear production becomes greater (figure \ref{GRTimeseries}(d)) or even exceeds the energy gain from meridional buoyancy flux (figure \ref{GRTimeseries}(g,h)). This result is also qualitatively consistent with \citet{heifetz2007generalized} (cf. their figure 7 bottom panels) for $\alpha=\beta=10$.
We observe that the increase in the proportion of energy gain from shear production at short time can be obtained through transition from the geostrophic to ageostrophic instability regime by increasing $\alpha$ (compare figs. \ref{GRTimeseries}(a,d)) or  increasing $\beta$ (compare progression in figs. \ref{GRTimeseries}(d,e) and in (b,e,h)). However, at long time, the energy gain is primarily drawn from the meridional buoyancy flux. 
Our analysis here shows that there is strong vertical momentum flux characteristic of ageostrophic modes at short-time $T<5$, in contrast with geostrophic instabilities for which the dominant meridional buoyancy flux that distributes the lateral buoyancy gradient horizontally. This difference in the energy pathways between geostrophic and ageostrophic modes is important because, as \citet{Boccaletti:07} noted, the ageostrophic instabilities could be significant due their timescales being less than the geostrophic adjustment timescale ($\mathcal{O}(f^{-1})$).\\

For 
unstable symmetric modes, the contribution of shear production to energy gain also increases with increasing $\beta$ value (compare figs. \ref{GRTimeseries}(c,f)). However, at longer time, contribution from buoyancy flux becomes equal to or exceeds the shear production (see figs. \ref{GRTimeseries}(c,f,i)). \citet{grisouard2018extraction} similarly showed a significant contribution from the energy drawn from the potential energy of the mean flow in two-dimensional numerical simulations of unstable symmetric modes. 
We will further explore this behavior in the next section using the optimal initial conditions for the growth in the finite amplitude regime.\\

\begin{figure}
     \includegraphics[width=1.\linewidth]{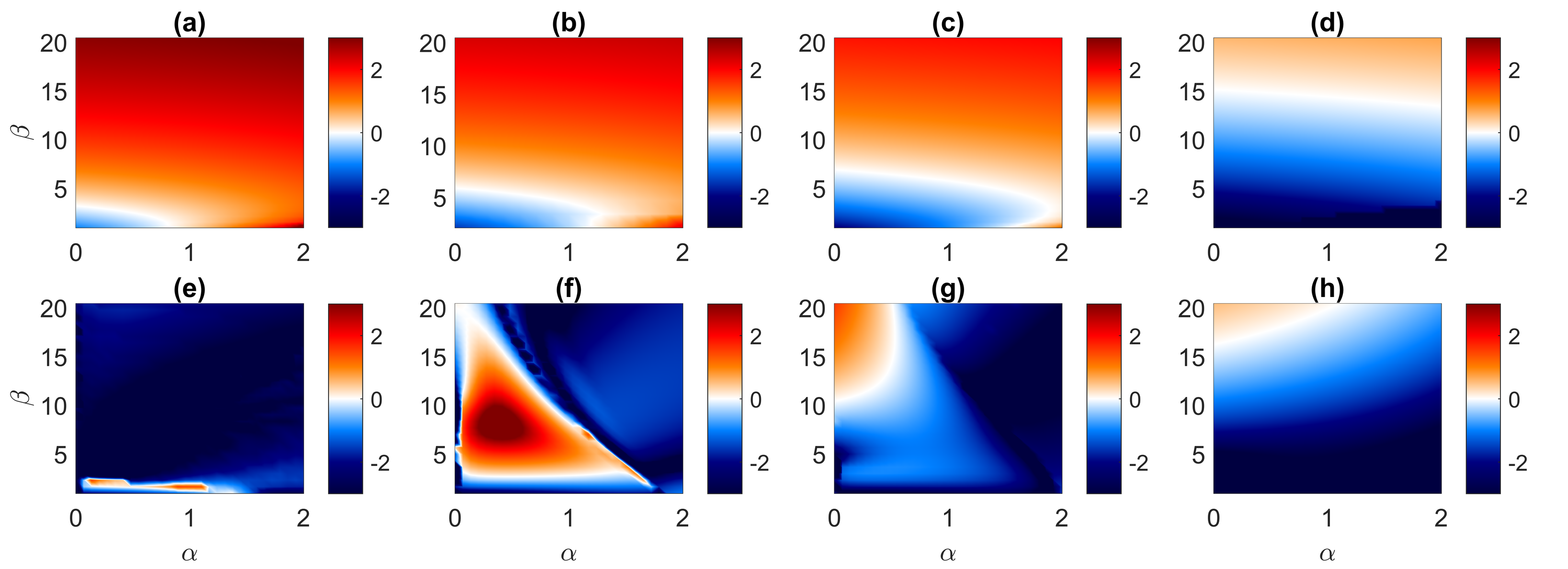}
        \caption{Parameter sweep over $\alpha=[0,2]$ and $\beta=[0.3,20]$ of log($\Gamma(\alpha,\beta)$), where $\Gamma$ is the ratio of shear production $-\tilde{u}\tilde{w}$ to meridional buoyancy flux $\tilde{v}\tilde{b}$ (top) at $T=0.5$ where transient growth occurs, (bottom) for normal growth: (a),(e) $\mbox{Ri}=2$, (b),(f) $\mbox{Ri}=0.92$, (c),(g) $\mbox{Ri}=0.5$, and (d),(h) $\mbox{Ri}=0.1$. }
        \label{GammaParam}
\end{figure}

As shown from the energy norm in (\ref{energy_norm}), we only need to consider the ratio of shear production to meridional buoyancy flux to characterize the exchanges in energy. Figure \ref{GammaParam} shows the log of the ratio, $\Gamma$, defined as
\begin{equation}
\Gamma = \frac{\int -\tilde{u}\tilde{w} dz}{\int \tilde{v}\tilde{b} dz},
\end{equation}
for various values of $\mbox{Ri}$ as functions of $\alpha$ and $\beta$. The top panel (figures \ref{GammaParam}(a-d)) shows log($\Gamma$) calculated at short times of optimization $T=0.5$,  whereas the bottom (figures \ref{GammaParam}(e-h)) shows log($\Gamma$) for normal growth. The patterns of ratio of shear production stress to buoyancy flux are strikingly different between normal mode and transient growth dynamics. Figure \ref{GammaParam}(g) matches to figure 2(a) in \citet{Stamper:17}.
For transient growth, this ratio is higher at larger $\mbox{Ri}$; as $\mbox{Ri}$ decreases, the region where the meridional buoyancy flux is greater than shear production  grows.  This progression can be explained by the transition from geostrophic to ageostrophic modes occurring at larger $\alpha$ as $\rm{Ri}$ decreases (see figure \ref{StoneValid}).\\

At optimization time $T=0.5$, for $\mbox{Ri}>1$ (figure \ref{GammaParam}(a)), the shear production is greater than the buoyancy flux for almost all values of $\alpha$ and $\beta$, except in a small region at low $\alpha$ and $\beta$ values, where the two energy fluxes are roughly equal or the meridional buoyancy flux is dominant. This corresponds to the region of higher normal mode growth rate in figure \ref{GammaParam}(a) due to geostrophic baroclinic instabilities. 
At intermediate $\mbox{Ri}$ values ($0.25 < \mbox{Ri} <1$), the buoyancy flux is dominant where both $\alpha$ and $\beta$ are small (figures \ref{GammaParam}(b-c)). Shear production becomes more dominant as 
unstable symmetric modes
become stronger (large $\beta$ values) and/or as ageostrophic instabilities emerge (large $\alpha$ values). 
At low $\mbox{Ri}$ values ($\mbox{Ri} \leq 0.25$), the region where the meridional buoyancy flux is dominant is larger than at higher $\mbox{Ri}$ values (figure \ref{GammaParam}(d)). Vertical stratification becomes weaker as $\mbox{Ri}$ decreases, while the horizontal buoyancy gradient that could drive the meridional buoyancy flux increases. Numerical simulations by \citet{arobone2015effects} similarly showed that near-symmetric unstable modes
generate flows that do not align with isopycnals and draw energy from the available potential energy reservoir of the background flow for several initial inertial periods.
However, as $\beta$ increases, 
unstable symmetric modes
become more dominant and $\Gamma$ becomes greater than $1$.\\

This analysis shows that  in the asymptotic limit as $t \rightarrow \infty$, the meridional transport overwhelms the shear production for a large portion of the $[\rm{Ri},\alpha,\beta]$ parameter space (see figs. \ref{GammaParam}(e-h)) with the exception of some mixed modes at $\rm{Ri}=0.92$ and
small regions of large-$\beta$ 
unstable symmetric modes
at low $\rm{Ri}$. 
However, we stress that it is not the case for the linear optimal perturbations, where for symmetric modes with large values of $\beta$ and ageostrophic modes with large $\alpha$ and/or $\beta$ values, the shear production is either equally important to or dominates the energetic dynamics. 

\subsection{Mechanisms of transient growth} 

\begin{figure}
     \includegraphics[width=1.\linewidth]{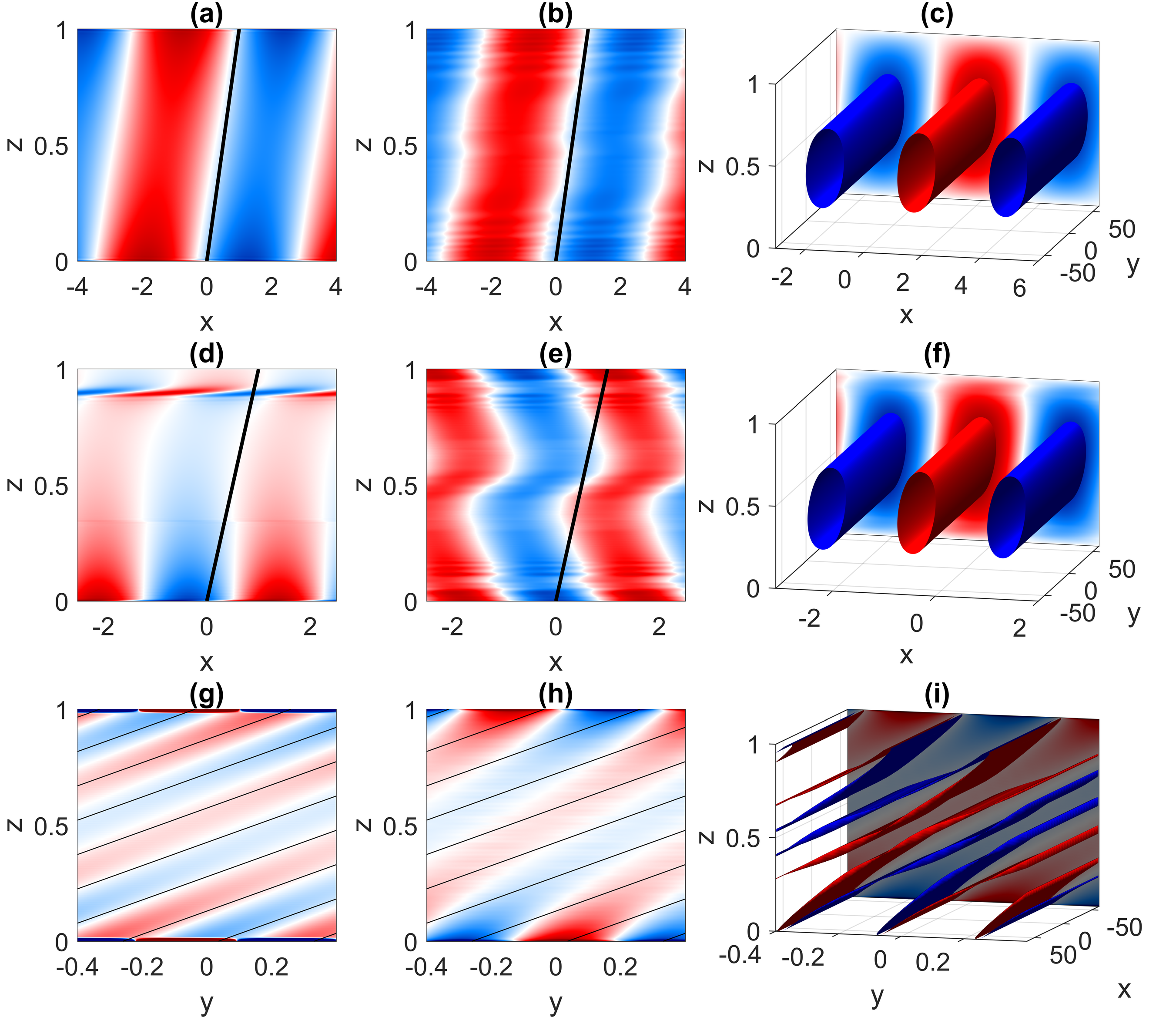}
        \caption{(Left) zonal-vertical buoyancy cross-sections of most unstable mode  (a,d) and meridional-vertical buoyancy cross-section of the first mode (stable) (g) ,(middle) zonal-vertical (b,e) and meridional-vertical (h) cross-sections of optimal perturbation initial condition for buoyancy, and (right) 3D structure of  optimal perturbation initial condition for vertical velocity. (a,b,c) are for geostrophic mode $[\alpha,\beta]=[1,0]$, (d,e,f) are for ageostrophic mode $[\alpha,\beta]=[1.8,0]$, (g,h,i) are for symmetric mode $[\alpha,\beta]=[0,10]$, which does not have any unstable modes. All values are for $\mbox{Ri}=2$ and $\delta=0.1$. Black lines in (a,b,d,e) represent the background zonal velocity $U(z)$ and black lines in (g,h) represent the background buoyancy $B(y,z)$.
        }
        \label{ICplot}
\end{figure}

In order to investigate the difference in amplification of growth rates by transient growth in geostrophic, ageostrophic and symmetric modes, we study the evolution of the  eigenvectors' profiles associated with the most unstable eigenmodes and optimal initial conditions that lead to optimal growth. As discussed in \S\ref{sec:Budget}, the shear production is a greater contributer to the energy budget than the meridional buoyancy flux at short-times for ageostrophic modes and symmetric modes. For this family of modes, the amplification of modal growth rates by transient dynamics is also greater than for geostrophic modes, for which meridional buoyancy flux is dominant over the shear production. In this section, we explore the mechanisms that are responsible for the differences in transient growth between each family of modes.\\

Figure \ref{ICplot} shows buoyancy cross-sections of strongly amplified representative unstable modes in the left panels, buoyancy cross-sections of initial conditions leading to optimal perturbation at $T=0.5$ in the middle panels, and three-dimensional structures of optimal initial conditions for vertical velocity in the right panels. All modes shown are for $\mbox{Ri}=2$ and $\delta=0.1$. The geostrophic mode with $[\alpha,\beta]=[1,0]$ has vortices that are uniform in the meridional direction with $z$-dependent vertical velocity perturbations (figure \ref{ICplot}(c)), and the buoyancy profiles for both the most unstable eigenmode and the optimal initial conditions that are aligned with the background shear,
indicated by black lines in figures \ref{ICplot}(a,b), respectively. 
We observe that both modal and optimal transient energy predominantly grows due to the meridional buoyancy flux (i.e. given by $vb$) for the geotrophic modes (see fig. \ref{GRTimeseries}(a)) and there is not as great of an amplification of growth rates by optimal perturbations (the transient growth rate is only twice the modal growth rate for this test case).\\

Buoyancy cross-sections and vertical velocity structure for ageostrophic mode with $[\alpha,\beta]=[1.8,0]$ are shown in figures \ref{ICplot}(d-f). The vertical velocity forms meridionally-uniform vortices similar to those of the geostrophic mode, but they are sheared near the vertical boundaries. The buoyancy profiles are significantly different between the most unstable mode and the optimal initial perturbation. Shear production (i.e. given by $-uw)$ has a non-zero contribution to the energy growth at short times for the optimal perturbations and the modal growth rates are substantially amplified by the optimal perturbations (the transient growth rate is $13$ times greater for this test case). \\

\begin{figure}
     \includegraphics[width=1.\linewidth]{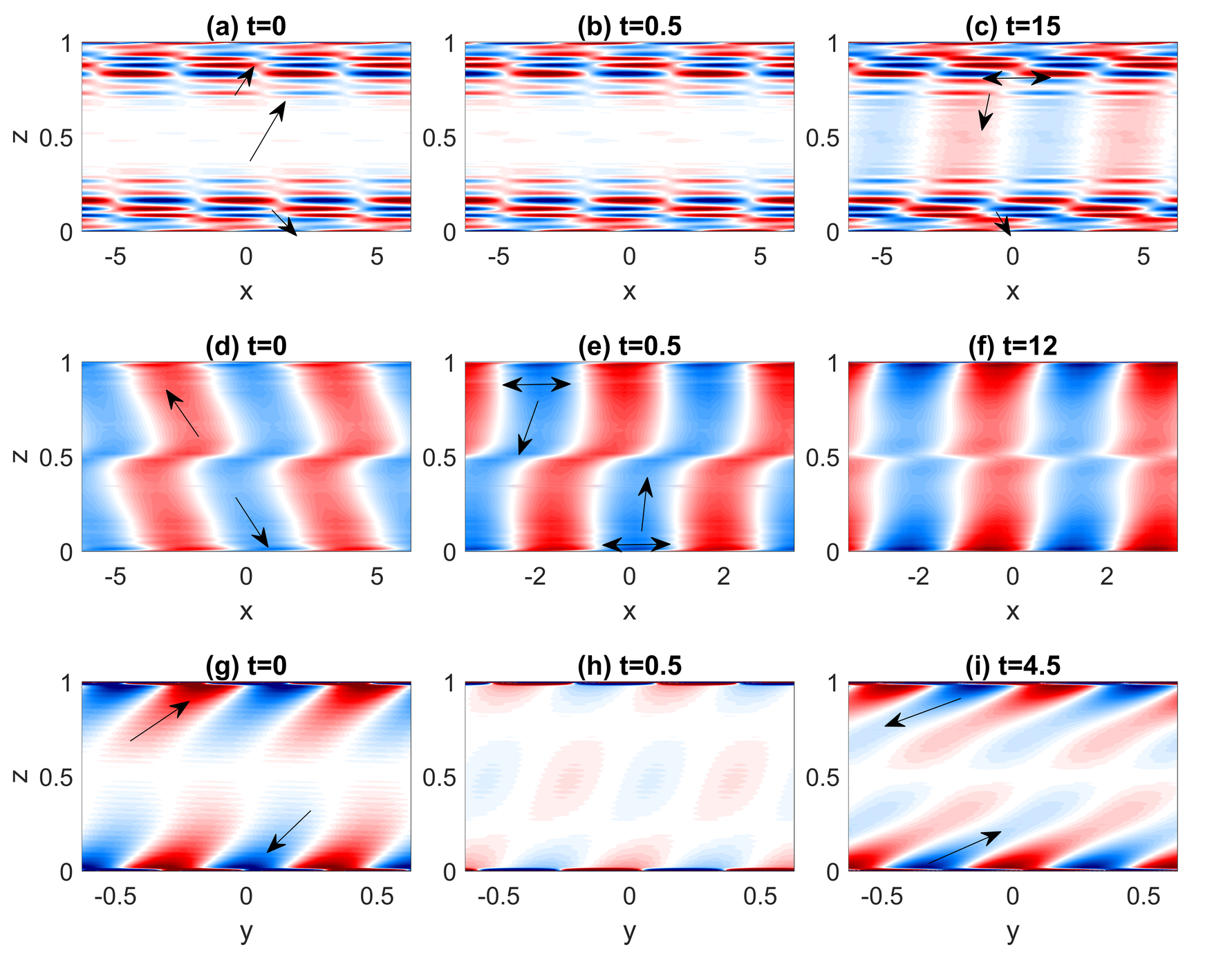}
     \caption{Temporal evolution of the difference between the optimal initial condition and the contribution of the unstable (or first) mode.  (a-c) geostrophic mode $\alpha=1$, $\beta=0$, $\mbox{Ri}=2$, $\delta=0.1$, (d-f) ageostrophic mode $\alpha=1.8$, $\beta=0$, $\mbox{Ri}=2$, $\delta=0.1$, and (g-i) symmetric mode $\alpha=0$, $\beta=10$, $\mbox{Ri}=2$, $\delta=0.1$. The arrows at the initial time show the direction of the perturbation travel toward the boundary or critical layer, and the arrows at later time show the reflection types (backward, forward, critical).}
     \label{ICevolution}
     \end{figure}

For a flow with a background thermal wind balance, wave-like instabilities can undergo critical reflection off horizontal surfaces and can lead to the irreversible energy exchange with the background flow \citep{grisouard2015critical,grisouard2016energy}. These reflections can be of three types, depending on the wave frequency: 1) backward reflection ($\omega < 1$), 2) critical reflection ($\omega=1$), and 3) forward reflection ($\omega>1$). Here we use wave frequency non-dimensionalized by $f$.  As noted by \citet{Nakamura:88}, critical layers can also serve as horizontal surfaces for the reflections of the instabilities in addition to the top and bottom boundaries.
In figure \ref{ICevolution}, the temporal evolution of the remainder between the optimal initial condition and the most unstable mode are shown to determine whether reflections, observed by \citet{grisouard2015critical,grisouard2016energy} occur for these instabilities. This residual perturbation was computed using the expression $\tilde{\mathbf{q}}(t)-\mathbf{q}^{\dagger,H}\tilde{\mathbf{q}}(t)\mathbf{q}$ where the adjoint mode $\mathbf{q}^{\dagger}$ is solution of the adjoint eigenvalue problem $\omega^H J \textbf{q}^\dagger = L^H\textbf{q}^\dagger$.\\

For baroclinic instabilities (geostrophic and ageostrophic), all of the retained modes, except the ones with nonzero growth rate, are purely imaginary and have frequencies $|\omega_i|\in [0,\alpha]$.  
For the geotrophic mode test case ($\mbox{Ri}=2, \alpha=1, \beta=0$), the retained waves have frequencies $|\omega_i| \leq 1$, meaning that we would expect backward and near-critical reflections, which we observe in figure 11(a-c). As shown by \citet{grisouard2016energy}, there is a net transfer of energy from the background flow to the perturbations through the exchange of potential energy as a result of these reflections, which is consistent with our finding of the meridional buoyancy flux primarily contributing to the transient energy growth for the geostrophic modes. Because there is not a significant difference between the optimal initial condition and the most unstable eigenmode, the energetic contributions and the wave reflections are small.   \\

The ageostrophic modes have larger $\alpha$, and thus, the frequencies of the retained modes span the regimes for the backward, critical and forward reflections. The contribution from the critical reflections are clearly shown in figure 11(d-f) for the ageostrophic mode test case ($\mbox{Ri}=2, \alpha=1.8, \beta=0$) through the intensification of the wave reflection contribution at the top and bottom boundaries and at the critical layers in the middle of the domain. The analysis of the energy budget by \citet{grisouard2016energy} showed that for higher wave frequencies in the forward reflection regime, there is transfer of kinetic energy from the background flow to the perturbations, which is consistent with the additional transient energy growth from the vertical shear production that we observe. Additionally, the critical and near-critical reflections have higher energy flux from the background flow \citep{grisouard2016energy}. The ageostrophic instabilities have modes that span a much wider range of frequencies than geostrophic modes with smaller $\alpha$ values (and frequencies are in the range $|\omega_i|\in [0,\alpha]$
), which allows the ageostrophic instabilities to potentially have more modes with the critical and near-critical reflections, explaining the greater energy growth rate amplification by the transient growth for ageostrophic modes compared with the geostrophic modes. \\ 


Finally, the test case for symmetric mode (figures \ref{ICplot}(g-i)) with $[\alpha,\beta]=[0,10]$ in the large $\mbox{Ri}$ regime is predicted to have zero energy growth rate by the asymptotic analysis by \citet{Stone:70}. We find that the buoyancy profile, as shown in figure \ref{ICplot}(g), and velocity profiles (not shown) for the first normal mode are aligned with the background stratification and shear. 
However, the optimal initial perturbation profile for buoyancy is steeper than the background stratification profile.
This mechanism relates to the fast-propagating modes interaction described in \citet{xu2007modal} and \citet{Xu:07}, who found that in the regime outside of the modal instability, the initial transient energy growth is driven by the interaction between the fastest propagating modes, whereas at a later time, the energy is generated by the slowest propagating modes and causes oscillations in energy (cf. figure \ref{GTimeseries}(d)).
In this case, cross-band circulation is tilted in the opposite direction of the surfaces of constant buoyancy, and the energy gain is driven by the perturbation buoyancy restoring force \citep{Xu:07}. 
Hence the remainder of the eigenspectrum does not appear to play a significant role in the transient growth process and the energy growth can be essentially associated with an inertia-gravity instability, resulting in the exchange of kinetic energy between the perturbation and the thermal wind shear through vertical momentum flux. This inertia-gravity instability is subject to critical reflection at the top and bottom boundaries. At a later time, the contribution to the energy growth from meridional buoyancy flux becomes dominant (cf. fig. \ref{GRTimeseries}(i)), corresponding to the backward reflection as seen in figure \ref{ICevolution}(i).
\citet{heifetz2008non} argued that the generalized growth of 
unstable symmetric modes
is relatively small. However, for large Richardson numbers, non-normal growth rates of 
unstable symmetric modes
are larger than the generalized growth rates of both baroclinic geostrophic and ageostrophic instabilities despite symmetric modes being asymptotically stable. In the case of fronts in the atmosphere, \citet{heifetz2008non} argued that such mechanisms can be found in the form of frontal lifting, or by orography. Analogously, in the ocean, the front could be forced at the lower boundary through internal tides originating from the bottom or wind stresses forcing the front through the surface.

\section{Three-dimensional simulations}\label{sec:3D}

In this last section, we investigate whether the linear transient growth dynamics are observed in the full nonlinear evolution, in particular whether we would observe the enhanced short-term energy growth predicted by the linear stability theory. We conduct DNS of a flow in a thermal wind balance initially perturbed by $[u,v,w,b]=[u(z),v(z),w(z),b(z)]e^{i\alpha x+i\beta y}$. The vertical profiles of the perturbations are either the most unstable mode or the linear optimal perturbation initial condition calculated for the optimization time $T=0.5$. We use three test cases at $\mbox{Ri}=2$ and $\delta=0.1$ for 1) geostrophic ($\alpha=1,\beta=0$), 2) ageostrophic ($\alpha=1.8,\beta=0$), and 3) symmetric ($\alpha=0,\beta=10$) instabilities, for which the initial perturbation conditions are shown in figure \ref{ICplot}. We additionally conduct simulations initialized with random vertical velocity and buoyancy profiles in order to evaluate whether the growing perturbations project onto the optimal modes, which is a more relevant case in the ocean or the atmosphere. The initial conditions of the optimal perturbations and random perturbations for $[u(z),v(z),w(z),b(z)]$ were normalized such that their maximum magnitudes matched those of the eigenfunction corresponding to the most unstable eigenmode for each instability type.\\

We solve the non-hydrostatic, incompressible Navier-Stokes equations in the Boussinesq limit, nondimensionalized using (\ref{adim}), which are equivalent to (\ref{Euler_ad}) with additional non-linear and viscous terms

\begin{subeqnarray}
\frac{D u}{D t}+ U\frac{\partial u}{\partial x} + w \frac{\partial U}{\partial z} -v + \rm{Ri} \frac{\partial p}{\partial x} - \frac{1}{\delta \rm{Re}} (\delta^2 u_{xx} + \delta^2 u_{yy} +  u_{zz}) &=&0 \\
\frac{D v}{D t}+ U  \frac{\partial v}{\partial x} +u + \rm{Ri} \frac{\partial p}{\partial y}-\frac{1}{\delta \rm{Re}} (\delta^2 v_{xx} + \delta^2 v_{yy} +  v_{zz})&=&0 \\
\delta^2 (\frac{D w}{D t} + U \frac{\partial w}{\partial x} )  - \rm{Ri} b + \rm{Ri}\frac{\partial p}{\partial z} -\frac{\delta}{\rm{Re}} (\delta^2 w_{xx} + \delta^2 w_{yy} + w_{zz}) &=&0 \\
\frac{D \hat{b}}{D t} + U\frac{\partial b}{\partial x} - \frac{v}{\rm{Ri}}   +w -\frac{1}{\delta \rm{Re} \rm{Sc}} (\delta^2 b_{xx} + \delta^2 b_{yy} +  b_{zz})  &=&0 \\
\frac{\partial u}{\partial x}  + \frac{\partial v}{\partial y}  +\frac{\partial w}{\partial z} &=&0
\label{NSDNS}
\end{subeqnarray}

In the remaining, we solve the transformed Navier-Stokes system (\ref{NSDNS}) using mixed finite differences in the vertical direction and pseudo spectral discretizations in both horizontal directions. Here, we implemented the same discretization as was previously used in \S \ref{sec:theory} for modal growth and in \S \ref{sec:TG} for transient growth in order to remain consistent. The above system is integrated in time with a semi-implicit formulation employing second-order backward Euler for the viscous terms while a second order Adams-Bashforth scheme is used for the advection terms \citep{passaggia2013adjoint}. The divergence-free velocity field is recovered using a standard projection method \citep{passaggia2014response} taking the divergence of the rescaled momentum equation (\ref{NSDNS}a-c) together with the divergence free condition (\ref{NSDNS}e). Dealiasing was performed applying a 1/2 rule on the nonlinear terms over both horizontal wavenumbers sets.\\

\begin{figure}
     \includegraphics[width=1.\linewidth]{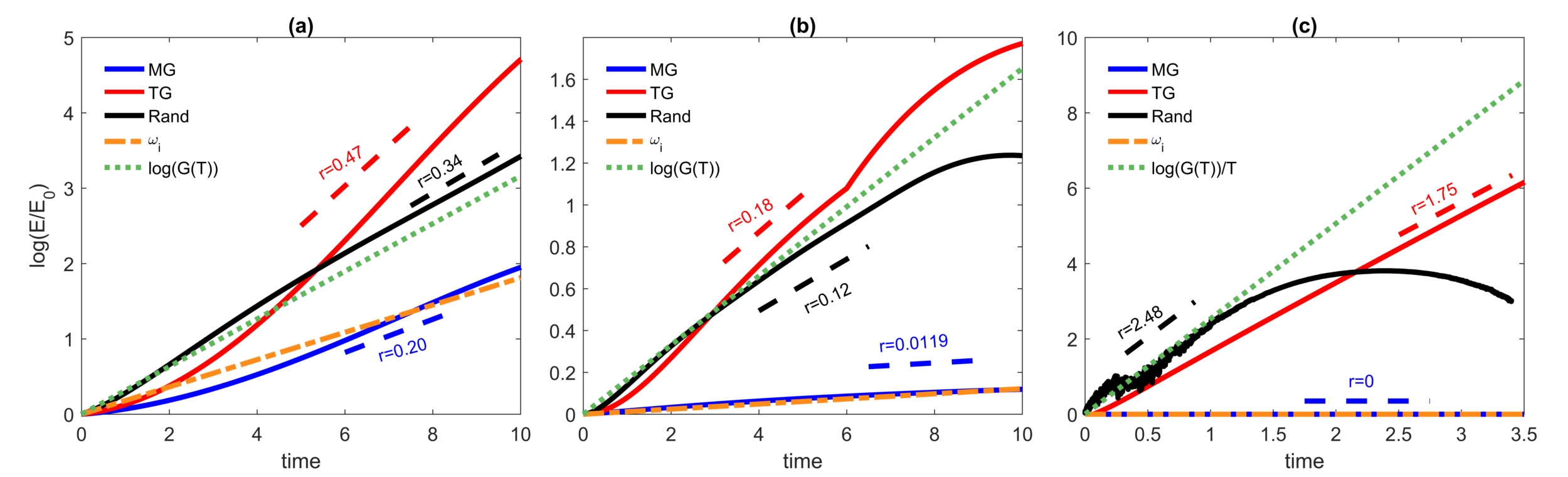}
        \caption{Energy growth over time for 3D simulations: (a) geostrophic mode ($\alpha=1,\beta=0$), (b) ageostrophic mode ($\alpha=1.8,\beta=0$), and (c) symmetric mode ($\alpha=0,\beta=10$). For each type of mode, $\rm{log}(E(t)/E_0)$ and average growth rate are shown for simulations initialized with most unstable mode ('MG' in blue), optimal perturbation ('TG' in red), and random initial condition ('Rand' in black). The modal growth rates ($\omega_i$) are shown with dash-dotted lines, and the transient growth rates at $T=0.5$ predicted from the linear stability theory ($\rm{log}(G(T))/T$) are shown with dotted lines. $r$ is the approximate energy growth rate compute from the slope at short time. All simulations are at $\mbox{Ri} = 2$.}
        \label{3DGrowth}
\end{figure}

We set the boundary conditions to be periodic in both zonal and meridional directions, and impose no-normal flow conditions at $z=[0,1]$, such that $w=0$ and $\partial u/\partial z = \partial v/\partial z =\partial b/\partial z = 0$. Note that the appropriate boundary condition for the buoyancy would be given by eq. (\ref{vort-eta-b}b) in order to strictly follow the inviscid problem. However, as shown later in the results section, the growth rates seem to be quite insensitive to this change in boundary condition. In addition, using eq. (\ref{vort-eta-b}b) at the solid boundary may lead to numerical instabilities near the boundaries in the case of the 3D DNS. 
We use $\nu=2 \times 10^{-6}$, which is the viscosity non-dimensionalized by $f/u_0^2$. Using $\delta=(H/L)/\rm{Ro}=0.1$, we compute $\rm{Re}=7\times 10^{6}$. We also set $\rm{Sc}=\nu/\kappa = 1$, where $\kappa$ is the diffusivity of the stratifying agent. For each simulation, we initialize the perturbation energy calculated using (\ref{energy}) to be 
$3\times10^{-5} E_B$, where $E_B  =\frac{1}{2} \int \int \int_V U^2 + \rm{Ri} B^2 dV$ is the energy of the background flow.
The code was validated against the stability analysis and results are shown in figure \ref{3DGrowth} for three different regimes where modal growth and transient growth results were accurately captured, at least for small amplitude initial conditions, that is initializing the code using the most amplified mode (denoted 'MG' in this section), the linear optimal perturbation (denoted 'TG' in this section), or white noise (denoted 'Rand' in this section).
The main question that we attempt to answer is the role of nonlinearities in transient growth and whether transient growth provides a faster mechanism than modal growth or random noise as an initial condition. Note that we are also looking at flow regimes that are strongly stratified ($\rm{Ri}=2$) which prevents the early rise of the secondary instabilities such as the ones reported by \citet{taylor2009equilibration,Taylor:11} for simulations initialized at $\rm{Ri}=0.5$. Similarly in their three-dimensional nonlinear evolutions, \citet{Stamper:17} did not observe a secondary energy growth for simulations initialized at $\rm{Ri}=1$, which were present for cases with initial $\rm{Ri}<1$.\\

The energy growth ($\rm{log}(E/E_0)$) over time for the 3D simulations is shown in figure \ref{3DGrowth} along with the predictions for the modal growth rate $\omega_i$ and transient energy growth rate at $T=0.5$ from the linear perturbation theory. The approximate energy growth rate $r$ for each simulation is computed from the slope.
The simulations with the 
unstable symmetric modes
were run for a shorter period of time, as these instabilities are typically faster growing than the baroclinic instabilities.
For all three types of modes (geostrophic, ageostrophic, and symmetric), the energy growth rate for the simulations initialized with the most unstable eigenmode (blue lines) is consistent with $\omega_i$. For the baroclinic instabilities (figure \ref{3DGrowth}(a-b)), the simulations initialized with the optimal perturbation profiles (red lines) have energy growth rates consistent with the transient growth rates at short time, but have some additional energy growth at a later time, possibly from the nonlinear interaction in the 3D simulations that are omitted in the linear stability analysis. We do, in fact, see three-dimensional perturbation flow structures develop even though the simulations are initialized with two-dimensional (non-varying in $y$) profiles.
However, for 
unstable symmetric modes
(figure \ref{3DGrowth}(c)), the growth rate for the 3D simulations is smaller than the predicted transient growth rate. 
Unstable symmetric modes
occur at smaller spatial scales than the baroclinic instabilities, such that viscosity may be important and dampens the energy growth.
Nonetheless, the energy growth rate is non-zero ($r=1.75)$, unlike the modal growth rate, suggesting that the 
symmetric modes
can grow in the regime of $\rm{Ri}>1$ given the optimal initial conditions. \\

The growth rate of the simulations initialized with a random profile (black lines in figure \ref{3DGrowth}) also have energy growth rates consistent with the predicted transient growth rates, at least at short time. For the geostrophic instabilities, the difference between the most unstable eigenvalue and the optimal transient growth rate is smaller than for the ageostrophic and the 
symmetric modes,
and the energy for 3D simulation continues to grow at the rate close to the optimal transient growth rate. However, the modal growth rates are significantly smaller for the ageostrophic ($\omega_i=0.012)$ and symmetric ($\omega_i=0)$ instabilities, and the energy growth rate is reduced at a later time approaching that of the modal growth rate. In the case of 
symmetric modes,
the energy begins to decrease at a later time, possibly due to viscous effects. This decay could also be a part of the oscillatory behavior in the energy growth, which is observed for the optimal energy growth in 
symmetric modes
at larger $\rm{Ri}$ (c.f figure \ref{GTimeseries}(d)). As discussed by \citet{Xu:07}, these oscillations are caused by the slowest propagating modes (the eigenmodes with purely real component with $|\omega_r|<1$).
In the time period shown here, the energy growth rate is non-zero for about two inertial periods, demonstrating that the 
symmetric modes
can occur for $\rm{Ri}>1$ for some period of time, when the initial perturbations are not exactly aligned with the background stratification, unlike the most unstable eigenmode. \\

\begin{figure}
     \includegraphics[width=1.\linewidth]{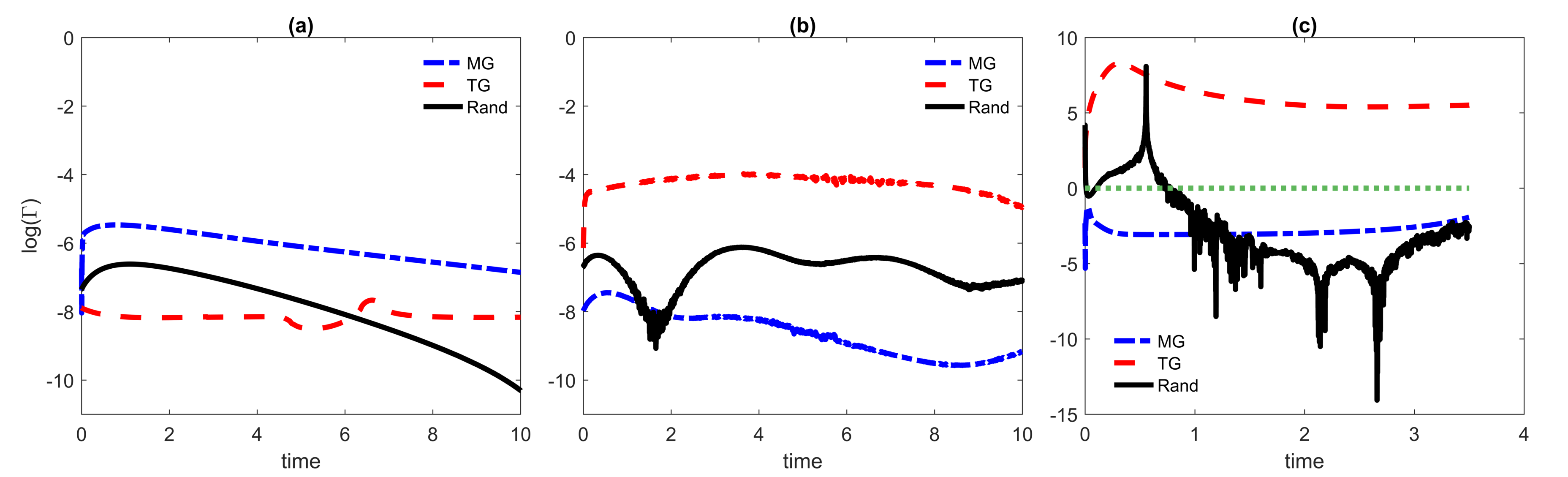}
        \caption{$\rm{log}(\Gamma)$ for each 3D simulation as a function of time: a) geostrophic mode ($\alpha=1,\beta=0$), (b) ageostrophic mode ($\alpha=1.8,\beta=0$), and (c) symmetric mode ($\alpha=0,\beta=10$). Notation as in fig. \ref{3DGrowth} with most unstable mode ('MG') in dot-dashed blue, optimal perturbation ('TG') in dashed red, and random initial condition ('Rand') in solid black. The green dotted line in (c) indicated zero.}
        \label{3DGamma}
\end{figure}

We further examine the contribution from the meridional buoyancy flux $vb$ and shear production $-uw$ to the energy growth by computing $\rm{log}(\Gamma)$ as a function of time for each simulation. These timeseries are shown in figure \ref{3DGamma} for the geostrophic modes (a), ageostrophic modes (b), and symmetric (c) modes. For both MG and TG simulations, $\rm{log}(\Gamma)$ is largely constant with time. Both the geostrophic and ageostrophic modes are predominantly driven by the meridional buoyancy flux, as predicted by the linear stability theory. However, for the geostrophic mode, the values of $\rm{log}(\Gamma)$ are closer between the simulations initialized by the most unstable mode and the optimal initial perturbation, suggesting that a similar mechanism drives the energy gain. In this regime, there is almost no energy gain from the shear production, so the more negative values of $\rm{log}(\Gamma)$ can be attributed to greater meridional buoyancy flux in the case of TG simulation. In the ageostrophic regime (fig. \ref{3DGamma}(b)), $\rm{log}(\Gamma)<0$ for all cases, but the TG simulation has $\rm{log}(\Gamma)$ has a higher value than that of the MG case. It is consistent with the prediction from the linear stability theory that a greater proportion of energy gain is drawn from the shear production in the case of the ageostrophic than of the geostrophic modes. Associated with this increase in $\rm{log}(\Gamma)$, the energy growth rate of the TG simulation of the ageostrophic mode is significantly greater than that of the MG simulation from this additional mechanism.\\

In the case of 
unstable symmetric modes
(fig. \ref{3DGamma}(c)), the TG case has $\rm{log}(\Gamma)>1$, consistent with strong shear production predicted by the linear stability theory. Note that while $\rm{log}(\Gamma)<1$ for the MG simulation, both the meridional buoyancy flux and the shear production are very small, as is evident from no energy growth rate in figure \ref{3DGrowth}(c). The simulation initialized with a random initial condition has initially ($t<1$) $\rm{log}(\Gamma)>1$, corresponding to the time period of large energy growth rate. For $t>1$, $\rm{log}(\Gamma)<1$, which corresponds to the time period when the energy growth approaching zero and then decaying. As mentioned before, \citet{farrell1988optimal} showed in a viscous Poiseuille flow, which is asymptotically stable, can experience rapid transient growth via the Orr mechanism, but the length of the time period for which this growth is sustained decreases with decreasing $\rm{Re}$. Thus, the decrease in energy growth rate that we observe in the random initial condition simulation for the symmetric case could be similarly due to the viscous energy dissipation, as the shear production that drives the energy gain is reduced.\\

\begin{figure}
     \includegraphics[width=1.\linewidth]{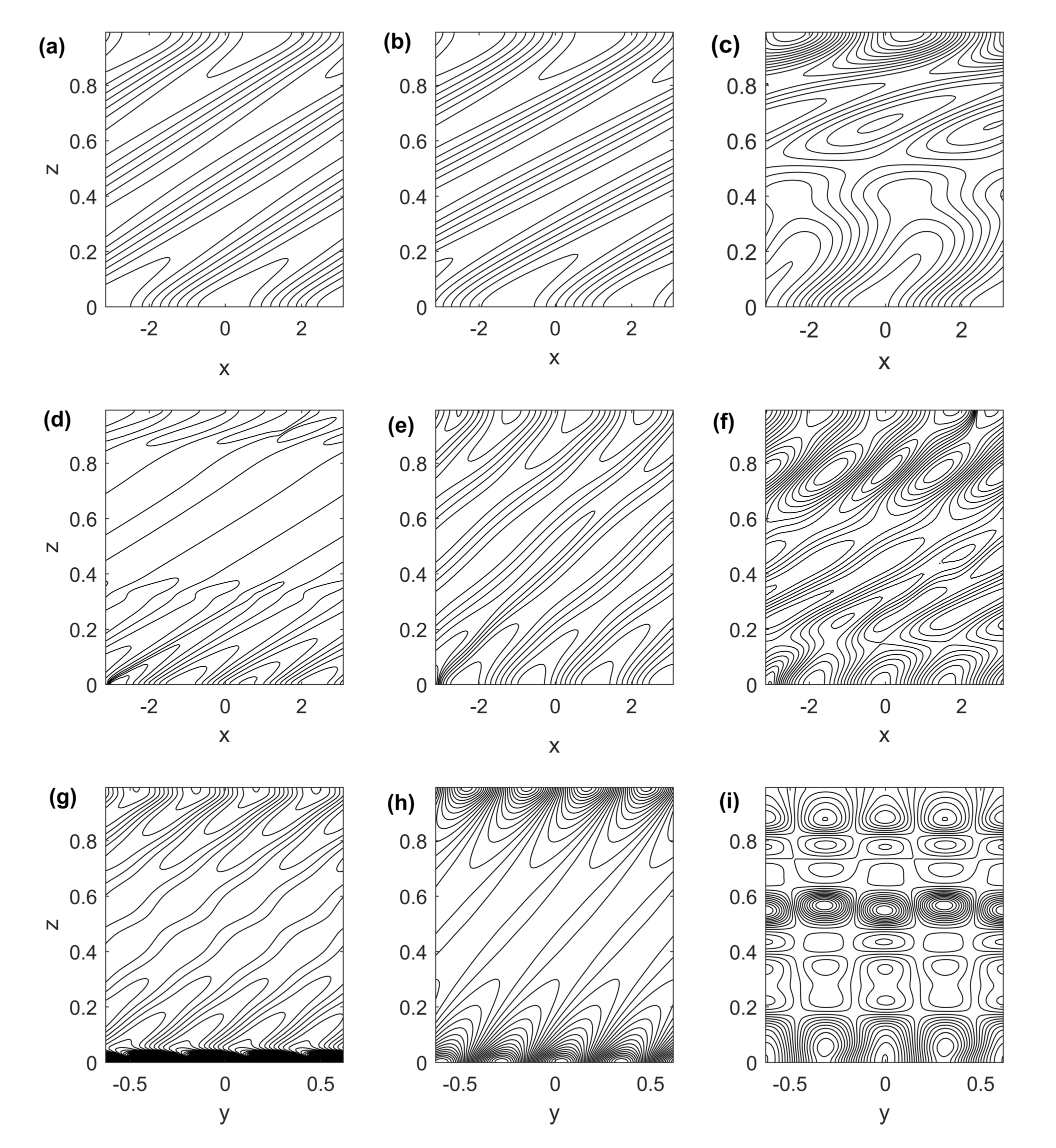}
        \caption{Perturbation buoyancy contours from 3D simulations at t=10 for the geostrophic (a-c) and ageostrophic (d-f) modes averaged over $y$ and at t=3.5 for the symmetric modes (g-i) averaged over $x$. Each simulation is initialized with: (left) most unstable mode, (middle) optimal perturbation, (right) random initial vertical profile.}
        \label{3DBuoCon}
\end{figure}

Figure \ref{3DBuoCon} shows the mean perturbation buoyancy contours at $t=10$ for the simulations initialized with the geostrophic modes (a-c) and the ageostrophic modes (d-f) and at $t=3.5$ for the symmetric modes (g-i). Buoyancy is averaged meridionally for the baroclinic modes, and zonally for the symmetric modes. The results for the simulations intialized with the most unstable eigenvectors are shown in left column (a,d,g), with the optimal perturbation profiles in the middle column (b,e,h), and with the random initial profiles in the right column (c,f,i). These buoyancy contours can be compared with the initial buoyancy distributions shown in figure \ref{ICplot} to examine the dominant mechanisms in each case. \\

In the case of  the baroclinic instabilities of both geostrophic and ageostrophic type, the energy growth drawn from the unstable eigenmode perturbations by tilting the isopycnals, as observed by comparing the initial conditions in figure \ref{ICplot}(a,d) with the final time contours in figure \ref{3DBuoCon}(a,d). This tilting is consistent with the baroclinic pathway predicted for these modes and can be expected from the backward reflection observed in figure \ref{ICevolution}(a-f). The isopycnals are similarly tilted for the simulations initialized with the optimal perturbations. In the case of geostrophic instabilities, the final time buoyancy contours of the modal and optimal growth simulations are very similar. The initial profiles also exhibited similar structure, and the growth rate amplification is not as large (a factor of $~2.5$) when the simulations are initialized with the optimal perturbation. The backward reflected waves, which drive the energy growth dynamics for the geostrophic mode, do not travel as far vertically compared with forward propagated waves and do not generate energy as much energy as waves with $\omega \geq 1$ \citep{grisouard2015critical}. \\

In the case of the ageostrophic instabilities, the optimal initial isopycnals (figure \ref{ICplot}(e)) were tilted against the background stratification unlike the modal initial buoyancy profile. In the final perturbation buoyancy profile, the perturbation isopycnals are more aligned with the background stratification, though the isopycnals are more tilted for the TG than the MG simulation. 
Although restratification is a second-order effect with respect to the perturbation amplitude, it is associated with
a greater amplification of the modal growth rate by the optimal perturbation in the case of ageostrophic instabilities than in the case of the geostrophic ones \citep{Boccaletti:07}. This amplification is due to the proportionally larger meridional buoyancy flux and the additional energy generation due to the perturbation buoyancy isopycnals being tilted against the background stratification, in response to the additional contribution from the forward and critical wave reflections (cf. fig. \ref{ICevolution}(d-f)).  
The energy generation from the shear production also contributes to the vertical transport resulting in a buoyancy field that is less stratified than the modal simulation, especially near the surface and bottom boundaries, shown in figure \ref{3DGamma}. The simulations initialized with the random profiles also undergo changes in stratification, and the isopycnal slopes at the final time approach those of the optimal perturbation cases, in particular in the upper part of the domain ($z>0.5$).\\

The buoyancy contours of the simulation initialized with the most unstable symmetric mode (fig. \ref{3DBuoCon}(g)) remain largely the same throughout the run period, as reflected by zero energy gain, with a small amount of mixing near the boundaries, possibly due to the non-linear dynamics. For the simulation initialized with the optimal perturbation (fig. \ref{3DBuoCon}(h)), the perturbation isopycnals are more vertical near the top and bottom boundaries, which is consistent with the critical reflections near the top and bottom boundaries (cf. fig. \ref{ICevolution}(h)) and the increased near-surface stratification in simulations with a non-zero horizontal density gradient of the slantwise convection by \citet{taylor2010buoyancy}.  
However, in the interior, the stratification is decreased compared to the initial conditions, indicative of strong vertical momentum ($-uw$) and buoyancy transport. 
The initial horizontal buoyancy gradient of the perturbation in this case is non-zero and the 
isovalues of buoyancy for the perturbation
are not aligned with the background horizontal gradient $B_y = -1/\mbox{Ri}$. As the isopycnals are restored to align with the background horizontal gradient and the modal solution, $M^4 = (b_y)^2$ is reduced and $N^2 = -b_z$ decreases to maintain $\mbox{Ri} = N^2 f^2/M^4$.
When the simulation was initialized with a random profile with 
unstable symmetric modes
(fig. \ref{3DBuoCon}(i)), the additional energy is extracted from the locally unstable horizontal gradients, and the buoyancy contours develop into more vertically-aligned vortices due to the vertical transport in contrast to the perturbation buoyancy contours that are more aligned with the background flow in the case of baroclinic instabilities (fig. \ref{3DBuoCon}(c,f)).

\section{Discussion}\label{sec:conclusion}

In this study, we compared the energy growth dynamics of the most unstable eigenmodes derived from the linear stability theory and of the optimal perturbations resulting from the non-normal interactions of all the eigenmodes. The results from the one-dimensional linear stability analysis were compared to the fully nonlinear three-dimensional simulations.
We examined a nonhydrostatic version of the Eady model of a fluid between two horizontal boundaries in a thermal wind balance. The energy growth rates were computed for the flow perturbed by baroclinic (geostrophic and ageostrophic) and 
unstable symmetric modes
over a range of zonal ($\alpha$) and meridional ($\beta$) wavenumbers, including the mixed-modes ($\alpha \neq 0$, $\beta \neq 0$). The transient energy growth rates were calculated using a singular value decomposition method that seeks optimal initial conditions to maximize the energy at a given target time. These calculations were validated by the transient growth rates approaching the most unstable eigenvalue (modal growth rate) as $t \rightarrow \infty$. The findings of the linear theory showed good agreement with the results of the nonlinear three-dimensional simulations. We obtained a good agreement between the simulations initialized with random perturbations profiles, which are the most relevant to the natural systems, and the linear theory, except for in the case of symmetric modes at large $\mbox{Ri}$, which needs to be explored in future work. \\

The main conclusions of this study are:
\begin{enumerate}
\item Transient energy growth rate at target time $T=0.5$ is greater than the modal energy growth for all combinations of $\alpha,\beta,\rm{Ri},$ and $\delta$. The target time of the convergence of the optimal transient growth rate to the modal growth rate is dependent on the instability type, and is generally faster for
the geostrophic instabilities. \\ 
\item The energy growth rates of the ageostrophic and 
symmetric modes
are more amplified by the transient dynamics (by one to two orders of magnitude) than the geostrophic instabilities (transient growth rates are greater than the modal growth rates by a factor of two to three). The energy growth can be non-zero at short time even for 
symmetric modes
at $\rm{Ri}>1$ in the regime that predicts no modal energy growth.
More striking is the fact that transient growth of these asymptotically stable  
symmetric modes
exceeds the growth of the most unstable modes of geostrophic instability type for $\mbox{Ri}\geq2$ meaning that transient growth is the dominant mechanism for strongly stratified fronts.\\
\item The transient energy growth is primarily driven by the wave reflections off the top and bottom domain boundaries and critical layers. The difference in the energy growth rate amplification of different instabilities is due to the presence of different types of reflections (backward, critical, forward), which depend on the frequencies of the eigenmodes and have different energy transfer rates with the background flow. Additionally, the energy of the 
symmetric modes,
in particular at $\rm{Ri}>1$, grows initially through the fast-propagating mode interaction mechanism and later through the 
slow-propagating mode interaction mechanism \citep{xu2007modal}, which can be triggered if the isopycnals of the perturbation are not aligned with the background meridional stratification.  \\
\item The magnitude of transient energy growth rate and the proportional contribution to the energy gain primarily increases with the increasing meridional wavenumber, even for mixed mode instabilities, for which both the meridional and zonal wave numbers are non-zero.\\
\item The transient energy growth rates are not significantly affected by the non-hydrostatic effects, unlike the modal growth rates that are reduced as $\delta$, which is proportional to the aspect ratio, increases.\\

\end{enumerate}

These findings have significant implications for the naturally occurring phenomena and can explain previous field observations. 
For instance, 
the successive order analysis of the subinertial mixed layer \citep{young1994subinertial}
suggests that the 
amplitude
of the ageostrophic 
effects
are small and may not act fast enough to restratify the surface layer before the next mixing event. 
Yet, as demostrated in this paper, the ageostrophic instabilities may in fact grow faster after the initial onset over several inertial periods and restratify the water column through the meridional buoyancy flux. Furthermore, 
the presence of the 
symmetric instabilities 
in the field observations is measured by the value of $\rm{Ri}$. Because we find that 
the asymptotically stable symmetric modes
can have transient growth rates even at $\rm{Ri}>1$, the potential contribution from 
these modes
could be more substantial and needs to be reassessed. \\



Interestingly, the optimal perturbation analysis predicts that the transient energy growth rates are mostly independent of the zonal wavenumber $\alpha$, and increase primarily with the meridional wavenumber $\beta$, meaning that 
an unstable symmetric mode
($\alpha =0$) and a mixed-mode three-dimensional instability ($\alpha \neq 0$) will have similar growth rates. We find that at higher $\beta$ values, the proportional energy gain from shear production, at least at short term, is greater, leading to larger rates of transient energy growth, even in cases where the meridional buoyancy flux is  dominant in the long-term dynamics.
This finding may explain the coupled effect of the symmetric and baroclinic instabilities: the 
unstable symmetric modes
occur at a faster time scale, and as they increase the stratification through the shear production, the mixed-mode instabilities are likely to be triggered. While at short term the mixed-mode instabilities continue to grow through shear production, they behave similarly to the baroclinic instabilities after a few inertial periods and restratify the fluid layer via horizontal buoyancy flux. In this framework, the energy growth rate may be sustained for many inertial periods, even though the modal analysis predicts a decrease in the energy gain in the transition between the symmetric and baroclinic instability regimes for $\rm{Ri}<1$. However, in this analysis, one must note that the transient growth rates are sensitive to the optimization target time \citep{xu2007modal}, and the results presented here may not be applicable to predicting the energy dynamics, including the dominant energy transfer mechanisms, several days after the onset of the instabilities, at which time, the modal analysis theory may be more appropriate. 
The temporal evolution of such coupled dynamics and the effect of the optimal initial perturbations will be explored in future work. \\

An important contribution of this work is the analysis of the full eigenspectrum of the modes obtained by solving the eigenvalue problem of the nonhydrostatic Eady model. While the nonhydrostatic component has been previously omitted from the linear stability analysis, the assumption of $H/L \ll 1$ may not hold true in many atmospheric \citep[e.g.]{nolan2017tornado} and oceanographic \citep{von2018observations} applications. 
Unlike the modal growth analysis, the transient growth calculations require the full eigenspectrum to be computed. We have identified the four different branches of the eigenspectrum and provided the selection criteria for the removal of the spurious modes that result from insufficient numerical resolution near critical layers. While these spurious modes can be eliminated by significantly increasing the vertical resolution because their growth rates decrease as the number of vertical discretization points increases, removing them from eigenspectrum a priori allows for a significantly faster computations of the transient growth rates. However, these relations only hold in a setup with a constant vertical background stratification, which is assumed by the Eady model, whereas the vertical buoyancy profiles found in nature may be more complicated \citep{Boccaletti:07,Thomas:13,ramachandran2018submesoscale}. The analysis of the eigenspectrum and the resulting transient growth rates for a generalized background stratification profiles will be addressed in future work.

\subsection*{Acknowledgements}
The authors acknowledge the support by the National Science Foundation Grant Number OCE-1155558 and OCE--1736989. We also wish to three anonymous referee for substantially improving the content of the paper and highlighting the possibilities for  reflections of near-inertial waves.

\bibliographystyle{apalike}
\bibliography{bib_jfm}

\begin{thebibliography}{}

\bibitem[Arobone and Sarkar, 2015]{arobone2015effects}
Arobone, E. and Sarkar, S. (2015).
\newblock Effects of three-dimensionality on instability and turbulence in a
  frontal zone.
\newblock {\em Journal of Fluid Mechanics}, 784:252--273.

\bibitem[Bakas and Farrell, 2009a]{bakas2009gravity_a}
Bakas, N.~A. and Farrell, B.~F. (2009a).
\newblock Gravity waves in a horizontal shear flow. part i: Growth mechanisms
  in the absence of potential vorticity perturbations.
\newblock {\em Journal of Physical Oceanography}, 39(3):481--496.

\bibitem[Bakas and Farrell, 2009b]{bakas2009gravity_b}
Bakas, N.~A. and Farrell, B.~F. (2009b).
\newblock Gravity waves in a horizontal shear flow. part ii: Interaction
  between gravity waves and potential vorticity perturbations.
\newblock {\em Journal of Physical Oceanography}, 39(3):497--511.

\bibitem[Boccaletti et~al., 2007]{Boccaletti:07}
Boccaletti, G., Ferrari, R., and Fox-Kemper, B. (2007).
\newblock Mixed layer instabilities and restratification.
\newblock {\em Journal of Physical Oceanography}, 37(9):2228--2250.

\bibitem[Brandt, 2014]{brandt2014lift}
Brandt, L. (2014).
\newblock The lift-up effect: the linear mechanism behind transition and
  turbulence in shear flows.
\newblock {\em Euro. J. Mech.-B/Fluids}, 47:80--96.

\bibitem[Brannigan et~al., 2017]{brannigan2017submesoscale}
Brannigan, L., Marshall, D.~P., Naveira~Garabato, A.~C., Nurser, A. J.~G., and
  Kaiser, J. (2017).
\newblock Submesoscale instabilities in mesoscale eddies.
\newblock {\em J. Phys. Ocean.}, 47(12):3061--3085.

\bibitem[Callies et~al., 2015]{callies2015seasonality}
Callies, J., Ferrari, R., Klymak, J.~M., and Gula, J. (2015).
\newblock Seasonality in submesoscale turbulence.
\newblock {\em Nature Comm.}, 6:6862.

\bibitem[Chandrasekhar, 1961]{Chandrasekhar:61}
Chandrasekhar, S. (1961).
\newblock Hydromagnetic and hydrodynamic stability.
\newblock {\em Clarendon, Oxford}.

\bibitem[Drobinski and Foster, 2003]{drobinski2003origin}
Drobinski, P. and Foster, R.~C. (2003).
\newblock On the origin of near-surface streaks in the neutrally-stratified
  planetary boundary layer.
\newblock {\em Boundary-layer meteorology}, 108(2):247--256.

\bibitem[Eady, 1949]{Eady:49}
Eady, E.~T. (1949).
\newblock Long waves and cyclone waves.
\newblock {\em Tellus}, 1(3):33--52.

\bibitem[Ellingsen and Palm, 1975]{ellingsen1975stability}
Ellingsen, T. and Palm, E. (1975).
\newblock Stability of linear flow.
\newblock {\em Phys. Fluids}, 18(4):487--488.

\bibitem[Ellis and Businger, 2010]{ellis2010helical}
Ellis, R. and Businger, S. (2010).
\newblock Helical circulations in the typhoon boundary layer.
\newblock {\em Journal of Geophysical Research: Atmospheres}, 115(D6).

\bibitem[Farrell, 1988]{farrell1988optimal}
Farrell, B.~F. (1988).
\newblock Optimal excitation of perturbations in viscous shear flow.
\newblock {\em The Physics of fluids}, 31(8):2093--2102.

\bibitem[Farrell and Ioannou, 1993a]{farrell1993optimal}
Farrell, B.~F. and Ioannou, P.~J. (1993a).
\newblock Optimal excitation of three-dimensional perturbations in viscous
  constant shear flow.
\newblock {\em Physics of Fluids A: Fluid Dynamics}, 5(6):1390--1400.

\bibitem[Farrell and Ioannou, 1993b]{farrell1993stochastic_a}
Farrell, B.~F. and Ioannou, P.~J. (1993b).
\newblock Stochastic dynamics of baroclinic waves.
\newblock {\em Journal of the atmospheric sciences}, 50(24):4044--4057.

\bibitem[Farrell and Ioannou, 1993c]{farrell1993stochastic_b}
Farrell, B.~F. and Ioannou, P.~J. (1993c).
\newblock Stochastic forcing of perturbation variance in unbounded shear and
  deformation flows.
\newblock {\em Journal of the atmospheric sciences}, 50(2):200--211.

\bibitem[Foster, 2013]{foster2013signature}
Foster, R. (2013).
\newblock Signature of large aspect ratio roll vortices in synthetic aperture
  radar images of tropical cyclones.
\newblock {\em Oceanography}, 26(2):58--67.

\bibitem[Gardner et~al., 1989]{Gardner:89}
Gardner, D.~R., Trogdon, S.~A., and Douglass, R.~W. (1989).
\newblock A modified tau spectral method that eliminates spurious eigenvalues.
\newblock {\em Journal of Computational Physics}, 80(1):137--167.

\bibitem[Gary and Helgason, 1970]{Gary:70}
Gary, J. and Helgason, R. (1970).
\newblock A matrix method for ordinary differential eigenvalue problems.
\newblock {\em Journal of Computational Physics}, 5(2):169--187.

\bibitem[Gnanadesikan et~al., 2005]{Gnanadesikan:05}
Gnanadesikan, A., Slater, R.~D., Swathi, P., and Vallis, G.~K. (2005).
\newblock The energetics of ocean heat transport.
\newblock {\em Journal of climate}, 18(14):2604--2616.

\bibitem[Grisouard, 2018]{grisouard2018extraction}
Grisouard, N. (2018).
\newblock Extraction of potential energy from geostrophic fronts by
  inertial--symmetric instabilities.
\newblock {\em Journal of Physical Oceanography}, 48(5):1033--1051.

\bibitem[Grisouard and Thomas, 2015]{grisouard2015critical}
Grisouard, N. and Thomas, L.~N. (2015).
\newblock Critical and near-critical reflections of near-inertial waves off the
  sea surface at ocean fronts.
\newblock {\em Journal of Fluid Mechanics}, 765:273--302.

\bibitem[Grisouard and Thomas, 2016]{grisouard2016energy}
Grisouard, N. and Thomas, L.~N. (2016).
\newblock Energy exchanges between density fronts and near-inertial waves
  reflecting off the ocean surface.
\newblock {\em Journal of Physical Oceanography}, 46(2):501--516.

\bibitem[Heifetz and Farrell, 2003]{Heifetz:03}
Heifetz, E. and Farrell, B. (2003).
\newblock Generalized stability of nongeostrophic baroclinic shear flow. part
  i: large richardson number regime.
\newblock {\em J. Atm. Sci.}, 60.

\bibitem[Heifetz and Farrell, 2007]{heifetz2007generalized}
Heifetz, E. and Farrell, B.~F. (2007).
\newblock Generalized stability of nongeostrophic baroclinic shear flow. part
  ii: Intermediate richardson number regime.
\newblock {\em J. Atmos. Sci.}, 64(12):4366--4382.

\bibitem[Heifetz and Farrell, 2008]{heifetz2008non}
Heifetz, E. and Farrell, B.~F. (2008).
\newblock Non-normal growth in symmetric shear flow.
\newblock {\em Quart. J. Royal Meteo. Soc.}, 134(635):1627--1633.

\bibitem[Lorenz, 1955]{Lorenz:55}
Lorenz, E.~N. (1955).
\newblock Available potential energy and the maintenance of the general
  circulation.
\newblock {\em Tellus}, 7(2):157--167.

\bibitem[Manning et~al., 2007]{Manning:07}
Manning, M.~L., Bamieh, B., and Carlson, J. (2007).
\newblock Descriptor approach for eliminating spurious eigenvalues in
  hydrodynamic equations.
\newblock {\em arXiv preprint arXiv:0705.1542}.

\bibitem[Molemaker et~al., 2005]{MolemakerM:05}
Molemaker, M.~J., McWilliams, J.~C., and Yavneh, I. (2005).
\newblock Baroclinic instability and loss of balance.
\newblock {\em J. Phys. Ocean.}, 35(9):1505--1517.

\bibitem[Morrison et~al., 2005]{morrison2005observational}
Morrison, I., Businger, S., Marks, F., Dodge, P., and Businger, J.~A. (2005).
\newblock An observational case for the prevalence of roll vortices in the
  hurricane boundary layer.
\newblock {\em Journal of the atmospheric sciences}, 62(8):2662--2673.

\bibitem[Nakamura, 1988]{Nakamura:88}
Nakamura, N. (1988).
\newblock Scale selection of baroclinic instability—effects of stratification
  and nongeostrophy.
\newblock {\em J. Atmos. Sci.}, 45(21):3253--3268.

\bibitem[Nolan et~al., 2017]{nolan2017tornado}
Nolan, D.~S., Dahl, N.~A., Bryan, G.~H., and Rotunno, R. (2017).
\newblock Tornado vortex structure, intensity, and surface wind gusts in
  large-eddy simulations with fully developed turbulence.
\newblock {\em Journal of the Atmospheric Sciences}, 74(5):1573--1597.

\bibitem[Omand et~al., 2015]{Omand:15}
Omand, M.~M., D’Asaro, E.~A., Lee, C.~M., Perry, M.~J., Briggs, N.,
  Cetini{\'c}, I., and Mahadevan, A. (2015).
\newblock Eddy-driven subduction exports particulate organic carbon from the
  spring bloom.
\newblock {\em Science}, 348(6231):222--225.

\bibitem[Orr, 1907]{Orr:07}
Orr, W. M.~F. (1907).
\newblock The stability or instability of the steady motions of a perfect
  liquid and of a viscous liquid. {P}art {I}: {A} perfect liquid.
\newblock {\em Proc. Royal Irish Acad. Sec. A: Math. Phys. Sciences}, 27:9--68.

\bibitem[Orszag, 1971]{Orszag:71}
Orszag, S.~A. (1971).
\newblock Accurate solution of the orr--sommerfeld stability equation.
\newblock {\em Journal of Fluid Mechanics}, 50(4):689--703.

\bibitem[Park et~al., 2017]{park2017instabilities}
Park, J., Billant, P., and Baik, J.-J. (2017).
\newblock Instabilities and transient growth of the stratified taylor--couette
  flow in a rayleigh-unstable regime.
\newblock {\em Journal of Fluid Mechanics}, 822:80--108.

\bibitem[Passaggia and Ehrenstein, 2013]{passaggia2013adjoint}
Passaggia, P.-Y. and Ehrenstein, U. (2013).
\newblock Adjoint based optimization and control of a separated boundary-layer
  flow.
\newblock {\em Eur. J. Mech.-B/Fluids}, 41:169--177.

\bibitem[Passaggia et~al., 2014]{passaggia2014response}
Passaggia, P.-Y., Meunier, P., and Le~Diz{\`e}s, S. (2014).
\newblock Response of a stratified boundary layer on a tilted wall to surface
  undulations.
\newblock {\em Journal of Fluid Mechanics}, 751:663--684.

\bibitem[Passaggia et~al., 2017]{Passaggia:17}
Passaggia, P.-Y., Scotti, A., and White, B. (2017).
\newblock Transition and turbulence in horizontal convection: linear stability
  analysis.
\newblock {\em J. Fluid Mech.}, 821:31--58.

\bibitem[Ramachandran et~al., 2018]{ramachandran2018submesoscale}
Ramachandran, S., Tandon, A., Mackinnon, J., Lucas, A.~J., Pinkel, R.,
  Waterhouse, A.~F., Nash, J., Shroyer, E., Mahadevan, A., Weller, R.~A.,
  et~al. (2018).
\newblock Submesoscale processes at shallow salinity fronts in the bay of
  bengal: Observations during the winter monsoon.
\newblock {\em Journal of Physical Oceanography}, 48(3):479--509.

\bibitem[Sarkar et~al., 2016]{sarkar2016interplay}
Sarkar, S., Pham, H.~T., Ramachandran, S., Nash, J.~D., Tandon, A., Buckley,
  J., Lotliker, A.~A., and Omand, M.~M. (2016).
\newblock The interplay between submesoscale instabilities and turbulence in
  the surface layer of the bay of bengal.
\newblock {\em Oceanography}, 29(2):146--157.

\bibitem[Schmid and Brandt, 2014]{Schmid:14}
Schmid, P.~J. and Brandt, L. (2014).
\newblock Analysis of fluid systems: Stability, receptivity, sensitivitylecture
  notes from the flow-nordita summer school on advanced instability methods for
  complex flows, stockholm, sweden, 2013.
\newblock {\em Applied Mechanics Reviews}, 66(2):024803.

\bibitem[Schmid and Henningson, 2012]{schmid2012stability}
Schmid, P.~J. and Henningson, D.~S. (2012).
\newblock {\em Stability and transition in shear flows}, volume 142.
\newblock Springer Science \& Business Media.

\bibitem[Scotti and Passaggia, 2019]{scotti2019diagnosing}
Scotti, A. and Passaggia, P.-Y. (2019).
\newblock Diagnosing diabatic effects on the available energy of stratified
  flows in inertial and non-inertial frames.
\newblock {\em J. Fluid Mech.}, 861:608--642.

\bibitem[Solberg, 1936]{Solberg:36}
Solberg, H. (1936).
\newblock Le mouvement d’inertie de l’atmosphere stable et son role dans la
  theorie des cyclones.
\newblock {\em Proces-Verbaux des s{\'e}ances de l’Union International de
  G{\'e}od{\'e}sie et G{\'e}ophysique (IUGG)}, pages 66--82.

\bibitem[Stamper and Taylor, 2017]{Stamper:17}
Stamper, M.~A. and Taylor, J.~R. (2017).
\newblock The transition from symmetric to baroclinic instability in the eady
  model.
\newblock {\em Ocean Dynamics}, 67(1):65--80.

\bibitem[Stone, 1966]{Stone:66}
Stone, P. (1966).
\newblock On non-geostrophic baroclinic stability.
\newblock {\em J. of Atm. Sciences}, 23.

\bibitem[Stone, 1970]{Stone:70}
Stone, P. (1970).
\newblock On non-geostrophic baroclinic stability: Part ii.
\newblock {\em J. of Atm. Sciences}, 27.

\bibitem[Stone, 1971]{Stone:71}
Stone, P. (1971).
\newblock Baroclinic stability under non-hydrostatic conditions.
\newblock {\em J. of Fluid Mech.}, 45.

\bibitem[Taylor and Ferrari, 2009]{taylor2009equilibration}
Taylor, J.~R. and Ferrari, R. (2009).
\newblock On the equilibration of a symmetrically unstable front via a
  secondary shear instability.
\newblock {\em Journal of Fluid Mechanics}, 622:103--113.

\bibitem[Taylor and Ferrari, 2010]{taylor2010buoyancy}
Taylor, J.~R. and Ferrari, R. (2010).
\newblock Buoyancy and wind-driven convection at mixed layer density fronts.
\newblock {\em Journal of Physical Oceanography}, 40(6):1222--1242.

\bibitem[Taylor and Ferrari, 2011]{Taylor:11}
Taylor, J.~R. and Ferrari, R. (2011).
\newblock Shutdown of turbulent convection as a new criterion for the onset of
  spring phytoplankton blooms.
\newblock {\em Limnology and Oceanography}, 56(6):2293--2307.

\bibitem[Thomas et~al., 2013]{Thomas:13}
Thomas, L.~N., Taylor, J.~R., Ferrari, R., and Joyce, T.~M. (2013).
\newblock Symmetric instability in the gulf stream.
\newblock {\em Deep Sea Res. Part II: Topical Studies in Oceanography},
  91:96--110.

\bibitem[Vallis, 2017]{vallis}
Vallis, G.~K. (2017).
\newblock {\em Atmospheric and oceanic fluid dynamics}.
\newblock Cambridge University Press.

\bibitem[Vasavada and Showman, 2005]{vasavada2005jovian}
Vasavada, A.~R. and Showman, A.~P. (2005).
\newblock Jovian atmospheric dynamics: An update after galileo and cassini.
\newblock {\em Reports on Progress in Physics}, 68(8):1935.

\bibitem[von Appen et~al., 2018]{von2018observations}
von Appen, W.-J., Wekerle, C., Hehemann, L., Schourup-Kristensen, V., Konrad,
  C., and Iversen, M.~H. (2018).
\newblock Observations of a submesoscale cyclonic filament in the marginal ice
  zone.
\newblock {\em Geophysical Research Letters}, 45(12):6141--6149.

\bibitem[Walters and Carey, 1983]{Walters:83}
Walters, R.~A. and Carey, G.~F. (1983).
\newblock Analysis of spurious oscillation modes for the shallow water and
  navier-stokes equations.
\newblock {\em Computers \& Fluids}, 11(1):51--68.

\bibitem[Wolfe et~al., 2008]{Wolfe:08}
Wolfe, C., Cessi, P., McClean, J., and Maltrud, M. (2008).
\newblock Vertical heat transport in eddying ocean models.
\newblock {\em Geophysical Research Letters}, 35(23).

\bibitem[Worsnop et~al., 2017]{worsnop2017gusts}
Worsnop, R.~P., Lundquist, J.~K., Bryan, G.~H., Damiani, R., and Musial, W.
  (2017).
\newblock Gusts and shear within hurricane eyewalls can exceed offshore wind
  turbine design standards.
\newblock {\em Geophysical Research Letters}, 44(12):6413--6420.

\bibitem[Xu, 2007]{xu2007modal}
Xu, Q. (2007).
\newblock Modal and nonmodal symmetric perturbations. part i: Completeness of
  normal modes and constructions of nonmodal solutions.
\newblock {\em J. Atmos. Sci.}, 64(6):1745--1763.

\bibitem[Xu et~al., 2007]{Xu:07}
Xu, Q., Lei, T., and Gao, S. (2007).
\newblock Modal and nonmodal symmetric perturbations. part ii: Nonmodal growths
  measured by total perturbation energy.
\newblock {\em Journal of the atmospheric sciences}, 64(6):1764--1781.

\bibitem[Young, 1994]{young1994subinertial}
Young, W. (1994).
\newblock The subinertial mixed layer approximation.
\newblock {\em Journal of physical oceanography}, 24(8):1812--1826.

\bibitem[Zemskova et~al., 2015]{Zemskova:15}
Zemskova, V.~E., White, B.~L., and Scotti, A. (2015).
\newblock Available potential energy and the general circulation: Partitioning
  wind, buoyancy forcing, and diapycnal mixing.
\newblock {\em Journal of Physical Oceanography}, 45(6):1510--1531.

\end{thebibliography}

\end{document}